\documentclass{acta}
\usepackage{supertabular,lscape,epsfig,siunitx,float}
\usepackage{amssymb}
\usepackage{amsmath}
\usepackage[symbol]{footmisc}

\SetPages{127}{158}
\SetVol{73}{2023}

\begin{document}
\newcommand{\TabApp}[2]{\begin{center}\parbox[t]{#1}{\centerline{
  {\bf Appendix}}
  \vskip2mm
  \centerline{ {\spaceskip 2pt plus 1pt minus 1pt T a b l e}
  \refstepcounter{table}\thetable}
  \vskip2mm
  \centerline{ #2}}
  \vskip3mm
\end{center}}

\newcommand{\TabCapp}[2]{\begin{center}\parbox[t]{#1}{\centerline{
  {\spaceskip 2pt plus 1pt minus 1pt T a b l e}
  \refstepcounter{table}\thetable}
  \vskip2mm
  \centerline{ #2}}
  \vskip3mm
\end{center}}

\newcommand{\TTabCap}[3]{\begin{center}\parbox[t]{#1}{\centerline{
   {\spaceskip 2pt plus 1pt minus 1pt T a b l e}
  \refstepcounter{table}\thetable}
  \vskip2mm
  \centerline{\footnotesize #2}
  \centerline{\footnotesize #3}}
  \vskip1mm
\end{center}}

\begin{Titlepage}
\Title{Candidates for Transiting Planets in OGLE-IV Galactic Bulge Fields}
\Author{M.~J.~~M~r~\'o~z$^1$,~~P.~~P~i~e~t~r~u~k~o~w~i~c~z$^1$,~~R.~~P~o~l~e~s~k~i$^1$,~~A.~~U~d~a~l~s~k~i$^1$,\\~~M.~K.~~S~z~y~m~a~\'n~s~k~i$^1$,~~M.~~G~r~o~m~a~d~z~k~i$^1$,~~K.~~U~l~a~c~z~y~k$^2$,~~S.~~K~o~z~{\l}~o~w~s~k~i$^1$,\\~~J.~~S~k~o~w~r~o~n$^1$,~~D.~M.~~S~k~o~w~r~o~n$^1$,~~I.~~S~o~s~z~y~\'n~s~k~i$^1$,~~P.~~M~r~\'o~z$^1$,\\~~M.~~R~a~t~a~j~c~z~a~k$^1$,~~K.~A.~~R~y~b~i~c~k~i$^{1,3}$,~~P.~~I~w~a~n~e~k$^1$, M.~~W~r~o~n~a$^1$}
{$^1$ Astronomical Observatory, University of Warsaw, Al. Ujazdowskie 4, 00-478 Warszawa, Poland\\
$^2$ Department of Physics, University of Warwick, Coventry CV4 7AL, UK\\
$^3$ Department of Particle Physics and Astrophysics, Weizmann Institute of Science, Rehovot 76100, Israel}
\Received{October 9, 2023}
\end{Titlepage}
\Abstract{
We present results of a search for transiting exoplanets in 10-yr long photometry with thousands of epochs taken in the direction of the Galactic bulge. This photometry was collected in the fourth phase of the Optical Gravitational Lensing
Experiment (OGLE-IV). Our search covered~$\approx 222 000$ stars brighter than
$I = 15.5~\mathrm{mag}$. Selected transits were verified using a probabilistic method. The search resulted in 99 high-probability candidates for transiting exoplanets. The estimated distances to
these targets are between $0.4~\mathrm{kpc}$ and $5.5~\mathrm{kpc}$, which is a significantly wider range than for previous transit searches. The
planets found are Jupiter-size, with the exception of one (named OGLE-TR-1003b) located in the hot Neptune desert. If the candidate is confirmed, it can be important for studies of highly irradiated intermediate-size planets.
The existing long-term, high-cadence photometry of our candidates increases the chances of detecting transit timing variations at long timescales. Selected candidates will be observed by the future NASA flagship mission, the Nancy Grace Roman Space Telescope, in its search for Galactic bulge microlensing events, which will further enhance the photometric coverage of these stars.}
{planetary systems -- Planets and satellites: detection -- Planets and satellites: fundamental parameters -- Techniques: photometric}

\section{Introduction}

Exoplanetary research is one of the most dynamically developing branches of astronomy. At the end of the twentieth century, advances in observing techniques enabled the detection of the first extrasolar planets. 
Currently, the multitude of exoplanetary discoveries is unprecedented, with the number of confirmed exoplanets exceeding 5500\footnote[1]{exoplanet.eu}. 
This large number of objects has allowed us to gain a broader understanding of the processes that govern planetary systems, their evolution, and their characteristics. Many studies have shown that planetary systems are much more diverse than our own Solar System
(see Zhu and Dong 2021 for a review).
\newline

Among the many methods for detecting exoplanets, the transit method plays a crucial role. It is based on observations of regular dimming of a star's brightness due to the presence of a planet blocking part of its light. The method was first used in 1999 to detect a previously known planet around the star HD 209458  (Henry \etal 2000, Charbonneau \etal 2000). 
The Optical Gravitational Lensing Experiment (OGLE) was the first survey to successfully apply the transit technique in discovering exoplanets. In observations of millions of stars of the Galactic disk, the project has found 219 planetary transit candidates (Udalski \etal 2002abc, 2008, Pont \etal 2008), of which seven were confirmed as planets (Konacki \etal 2003ab, Pont \etal 2004, 2008, Bouchy \etal 2004, Udalski \etal 2008). By announcing the first exoplanet discovered using the transit technique and confirmed through radial velocity measurements -- OGLE-TR-56b, OGLE paved the way for extremely fruitful ground-based surveys for transiting exoplanets, such as WASP (Pollacco \etal 2006), HATNet (Bakos \etal 2004), NGTS (West \etal 2016) and space missions, like Kepler (Borucki \etal 2010) and TESS (Ricker \etal 2015).
However, the majority of exoplanetary surveys are limited to bright nearby stars. Only the Kepler mission was capable of discovering transiting planets around distant stars, up to 1 kpc away. This ability led to significant differences when compared to results from surveys with a smaller distance range, such as those using the radial velocity method (which primarily detects planets at distances up to 100 pc). In particular, there is a difference in the occurrence rate of hot Jupiters orbiting nearby stars (Cumming \etal 2008, Mayor \etal 2011) and distant stars (Howard \etal 2012, Guo \etal 2017).
Discovering transiting planets around faint stars in microlensing survey data offers a unique opportunity to investigate short orbital period planets across a wide range of galactocentric distances.
\newline

Over its more than three decades of operation, the OGLE survey has had a profound impact on many fields of modern astrophysics. In this study, we present a large sample of high-probability transiting planet candidates selected from the OGLE survey database.  
This database has not been searched for planetary transits since the early 2000s. 
The analyzed data were collected between 2010 and 2019 and consist of between 11 and 17 thousand measurements per object. Such long, regular, high-quality, and homogeneous observations have not been matched by any other survey capable of detecting transiting exoplanets.
\newline

In Section 2, we provide a brief description of the OGLE survey and the analyzed photometric observations. Section 3 introduces the steps taken to find, verify, and characterize planetary candidates. In Section 4, we present the final catalog of candidates. The discussion in Section 6 includes our thoughts on the future of our targets. Finally, in Section 7, we summarize our work.

\section{The OGLE survey}
The primary objective of the OGLE survey is a continuous monitoring of billions of stars within the Galactic bulge, disk, and the Magellanic Clouds to study their photometric variability. For this purpose, the survey employs a 32-detector mosaic camera attached to the 1.3-m Warsaw telescope at the Las Campanas Observatory, Chile, which is operated by the Carnegie Institution for Science.  The installation of this camera in March 2010 marked the commencement of the OGLE's fourth phase (Udalski \etal 2015). The camera has a field of view of 1.4 deg$^2$ and a resolution of 0.26 arcsec/px.
The exceptional observing conditions at Las Campanas allow OGLE to collect photometric data from some of the most densely populated regions of the sky with very high precision.
For our analysis, we used data from ten observational seasons, between 2010--2019. Due to the COVID-19 pandemic restrictions, OGLE temporarily stopped its regular operations in March 2020, resuming observations in August 2022.  The majority of observations  were taken through the Cousins $I$ filter, with additional data collected using the Johnson $V$ filter. For our studies, we chose four of the most frequently observed fields in the direction of the Galactic bulge, with up to 16~800 data points per object and a cadence as short as 19~min (see Table 1). The exposure times of those observations were 100~s in $I$ and 150~s in $V$. The photometric data was reduced using the Difference Image Analysis (DIA,  Wo{\'z}niak 2000), a technique developed specifically for dense stellar fields  (Alard and Lupton 1998). 

\begin{table}[h!]  
\centering
\medskip
\TabCapp{12pt}{OGLE-IV fields searched for transiting planets}
\begin{tabular}{ccc}
\hline \noalign{\smallskip}
OGLE field  & $N_{\rm{obs}}$ &  $N_{*} (I \leq 15.5,~(V-I) \leq 2 ) $\\\noalign{\smallskip}
\hline \noalign{\smallskip}
BLG501 &  15 700 & 29 600\\
BLG504 &  11 100 & 23 200\\
BLG505 &  16 800  & 52 600\\
BLG512 &  14 500 & 116 700\\\noalign{\smallskip}
\hline \noalign{\smallskip}
SUM &  & 222 100\\\noalign{\smallskip}
\hline
\end{tabular}

    \label{tab:my_label}
\end{table}

We focused our search on bright main-sequence stars in the Galactic disk by employing the following criteria for brightness and color selection: $I \leqslant $15.5~mag and $(V-I) \leqslant  2~$mag. This is illustrated in Fig. 1. In total, approximately 222~100 objects within the analyzed fields met these criteria (Table 1). After reviewing the light curve of the known planetary transit OGLE-TR-10 (Udalski \etal 2002a) detrended with various methods, we opted for a straightforward correction method using the median value from each season. This corrects, in a sufficient way, the decrease in brightness, caused by the proper motions shifts of targets with respect to the reference images. To minimize the effects of cosmic rays, weather conditions, bad CCD columns, satellite and asteroid flybys, and other factors, we implemented a sigma clipping technique, excluding data points exceeding $2\sigma$ above the normalized flux.
\begin{figure}[h!]
\begin{center}
\includegraphics[width=0.9\textwidth]{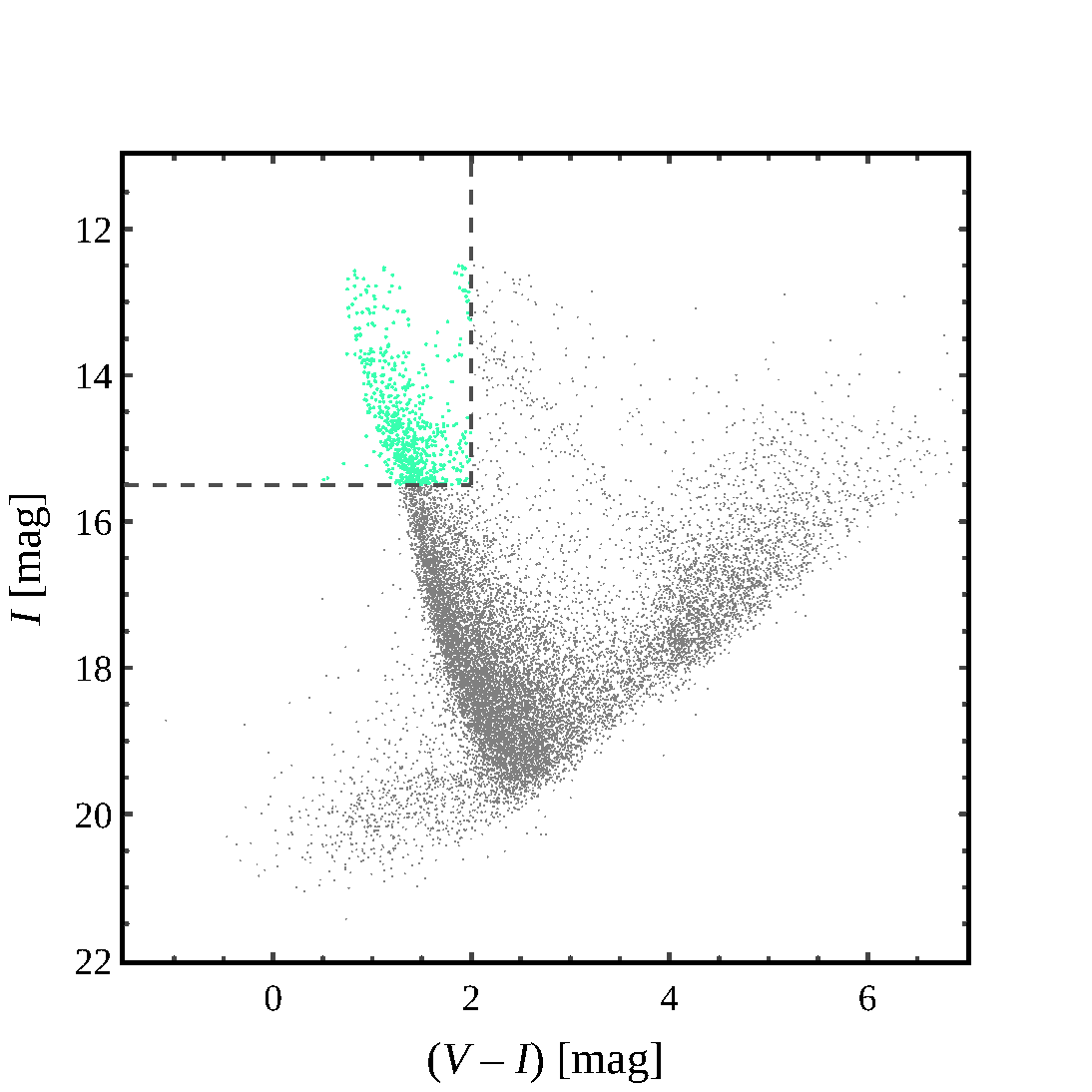}
\label{fig:CMD}  
\end{center}

\FigCap{Color-magnitude diagram for the OGLE sub-field BLG501.15, with the marked selection cuts.}
\end{figure}

\section{Transit Detection}
\subsection{Initial Transits Search}
A commonly used technique for transit searches involves the least squares
fitting of transit templates to the light curves, folded according to a grid of trial periods. In our work, we applied the Transit Least Squares (\textsf{TLS}) algorithm (Hippke and Heller 2019). This modern method implements a more realistic transit shape compared to the rectangular template used in the traditional Box Least Squares (\textsf{BLS}) algorithm (Kov{\'a}cs \etal 2002). By including limb darkening and the planetary ingress and egress effects in the shape of the light curve, \textsf{TLS} is more sensitive to low signal-to-noise ratio (SNR) transits. The \textsf{TLS} template is based on the transit model of Mandel and Agol (2002), with parameters optimized based on a set of previously known exoplanet transit detections. The use of realistic templates increases the detection efficiency by 5-10 percent at the expense of computational speed (Hippke and Heller~2019). 

Ground-based observations, such as the ones collected by the OGLE project, present additional challenges in the detection of periodic signals compared to continuous data collected by space telescopes.
Various natural phenomena, like the day-night cycle, moonlight contamination, etc., can produce periodic changes in the structure of observational data (\eg Baluev 2012). 
To minimize false positive signals caused by these phenomena, we added an option to  \textsf{TLS} to exclude problematic sections of tested periods from the period grid. We conducted a trial search to determine which periods should be excluded and which \textsf{TLS} parameters would be most suitable for our purposes. We initiated the search by excluding the most prominent multiples of the sidereal day. Subsequently, throughout this trial run, we extended the excluded period ranges to minimize recurring false positive signals. This increased the sensitivity to transit signals and reduced the \textsf{TLS} calculation time (final excluded period ranges are listed in  Table 2).  After the preliminary run, we fixed the following parameters: oversampling factor = 4, transit duration grid step = 1.1, minimum period = 0.6~d, and maximum period = 500~d. For the definition of the parameters and other details of the \textsf{TLS}, we refer to Hippke and Heller~(2019).

\begin{table}[h!]  
\centering
\TabCapp{12pt}{Period ranges excluded from the \textsf{TLS} search grid.}
     \begin{tabular}{r@{ -- }l@{}c}
 \hline\noalign{\smallskip} \multicolumn{2}{c}{Excluded period range}  
 &  Comment \\\noalign{\smallskip}
\hline
\noalign{\smallskip}
0.6640&0.6665 & 2/3 sidereal day \\
0.7474&0.7475 & 3/4 sidereal day \\
0.990&1.010 & 1 sidereal day \\
1.329 &  1.331 & 4/3 sidereal day \\
1.493 & 1.499 & 3/2 sidereal day \\
1.661 & 1.664 & 5/3 sidereal day \\
1.988 & 2.001 & 2 sidereal day \\
2.492 &  2.497 & 5/2 sidereal day \\
2.981 & 3.002 & 3 sidereal day  \\
3.988 & 3.9905 & 4 sidereal day \\
3.4895 &  3.490 & 7/2 sidereal day \\
4.985 &  4.9875 & 5 sidereal day \\
5.9825 & 5.984 & 6 sidereal day \\
6.980 & 6.982 & 7 sidereal day\\
54.678 &  55.180 & unknown\\
58.603 & 58.604 & unknown\\
64.929 & 65.931 & unknown\\
68.410 & 68.412 & unknown\\
100.058 & 100.062 & unknown\\
115.205 &  115.210 & unknown\\
146.130 & 146.140 & unknown\\
\noalign{\smallskip}
\hline
\end{tabular}
    \label{tab:periods}
\end{table} 
Due to residual contamination, we decided to visually inspect the results of the \textsf{TLS} calculations for almost all the objects. We automatically excluded only signals caused by breaks in the folded data or no signal detections. 
True signals were classified into the following groups: transits, sinusoidal variations, and other variability.
In the next step, we double-checked all the signals classified as transits to exclude obvious eclipsing binary systems (based on transit depth and the presence of a secondary eclipse). After this process, we were left with 4601 potential planetary-like transits.

\subsection{Probabilistic Evaluation}
The major disadvantage of the transit method is the need for additional confirmation of the planetary nature of transiting objects. Firstly, from the shape of the transit, we can only deduce the radius of the transiting object. Therefore, we cannot distinguish between hot Jupiters, brown dwarfs, and low-mass M-type stars, as they can all have similar sizes. Secondly, additional light blended in the photometric aperture can shallow the transit depth and lead to underestimated radius measurements. The traditional way of resolving this degeneracy is to measure the mass through radial velocity (RV) observations. Nevertheless, there are indirect methods of validation in the absence of RV measurements. One of them is the Validation of Exoplanet Signals using a Probabilistic Algorithm (\textsf{VESPA}, Morton 2012). \textsf{VESPA} is a widely used and well-established publicly available software package that performs statistical validation. We used it to evaluate 4601 candidates against astrophysical false positive scenarios. By applying a simple trapezoidal transit model, known parameters of the host star, and assumptions about the populations and distribution of field stars, the algorithm calculates the probability of the analyzed object being a planet given the observed signal $\rm{Pr(planet | signal)}$. 
\textsf{VESPA} considers the following false positive scenarios:
    \begin{itemize}
        \item  blended eclipsing binary (BEB), in the case of non-associated eclipsing binary systems blended within the photometric aperture of the target star,
        \item hierarchical eclipsing binary (HEB), in case of the target is a hierarchical triple system in which two components are eclipsing,
        \item eclipsing binary (EB). 
    \end{itemize}
For detailed descriptions, we refer to Morton (2012, 2016).

We prepared normalized flux light curves in the same way as for the \textsf{TLS}. As initial values required by \textsf{VESPA} (the orbital period $P$, planet-to-star radius ratio $R_{\rm{p}}/R_*$), we used values fitted by the \textsf{TLS}.  We derived the maximum allowed depth of a potential secondary eclipse for each object by searching the phase-folded light curve for the deepest signal at any phase outside of the primary transit (Morton 2016). We set the maximum angular distance (parameter \textit{maxrad} in \textsf{VESPA}) from the target star where a potential blending star might be, as $2\zdot\arcs6$, which is 10 times the pixel size of the OGLE-IV camera or roughly twice the average seeing of the images.

The obtained values of probability for the planetary origin of transits are shown in the histogram in Fig. 2. For further analysis, we decided to use only the cases for which the probability is greater than $80\%.$ 
Scenarios with the highest probability for each analyzed transit are shown in Fig. 3. 
For 1110 transits, \textsf{VESPA} calculation failed, in most cases due to not converging transit model.

\begin{figure}[h!]
    \includegraphics[width=0.99\textwidth]{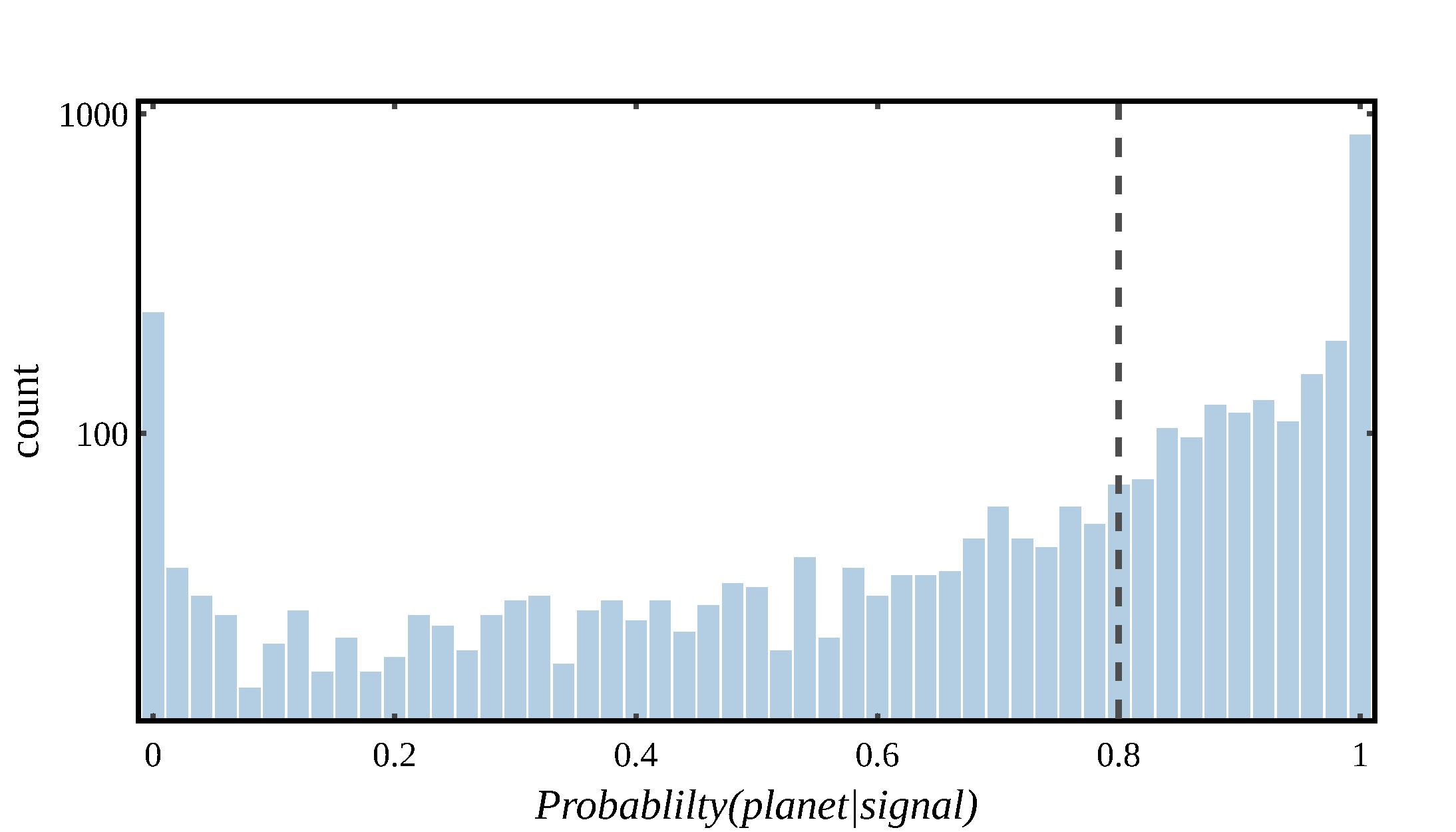}
    \FigCap{Distribution of $\rm{{Pr}(planet | signal)}$ values for 3491 potential planetary-like transit. The dashed line marks the minimum value of the probability for which transit was used in further analyzes.}{}
\end{figure}
\begin{figure}[h!]
    \includegraphics[width=0.99\textwidth]{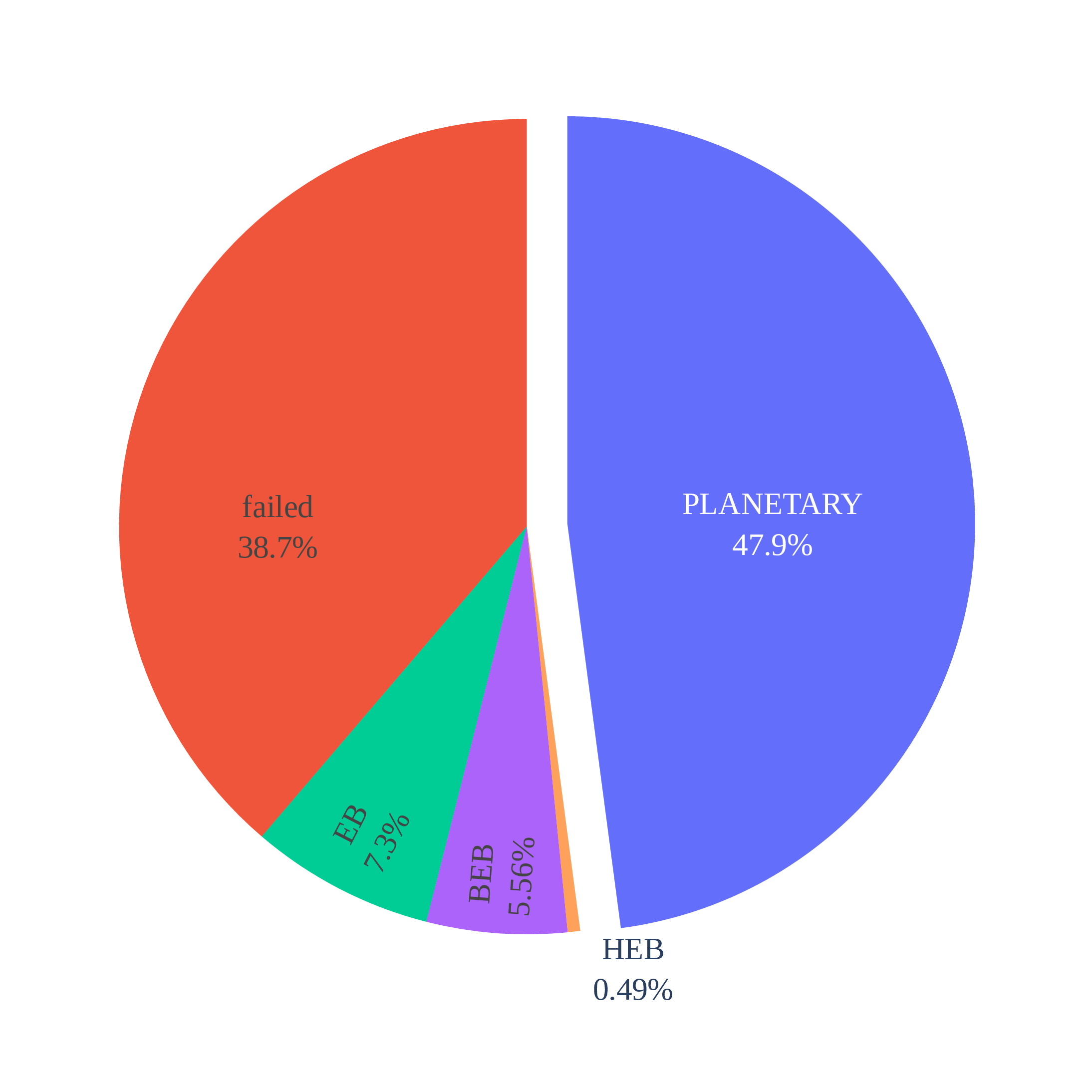}
    \FigCap{Most probable scenario for 4601 potential planetary-like transits. Each slide represents a different scenario simulated with \textsf{VESPA}: BEB -- blended eclipsing binary, HEB -- hierarchical triple system with eclipse, EB -- eclipsing binary. PLANETARY -- planetary transits, failed -- \textsf{VESPA} calculations failed. }{}
\end{figure}

\subsection{Modeling the Candidates}
We chose \textsf{EXOFASTv2} (Eastman \etal 2019) for modeling our candidates, which is Markov Chain Monte Carlo (\textsf{MCMC}) transit modeling code. To properly model the transit, it is essential to model both the host star and the planet. \textsf{EXOFASTv2} offers the unique capability of simultaneously constraining host star parameters using both the transit shape and stellar models, employing isochrones from stellar evolutionary tracks and spectral energy distribution (SED). These features of the code are particularly helpful in constraining system parameters, especially when spectroscopic observations are unavailable.

For modeling, we used OGLE observations in both the $I$- and $V$-bands, which were detrended and normalized as in previous steps. When an object was observed in multiple OGLE fields, we added those observations as separate data sets. In a few cases, this notably enlarged the number of observations, as for example for OGLE-TR-1051 (see Fig. 10).  We converted the original OGLE time stamps from Heliocentric Julian Date (HJD) to Barycentric Julian Date ($\rm{BJD_{TDB}}$) using IDL codes written by Eastman (2010). As starting values for the period $P_{\rm{orb}}$ and the time of conjunction $T_{\rm{C}}$, we implemented values from the \textsf{TLS} output. Whenever available, we incorporated parallax $\varpi$ and host effective temperature $T_{\rm{eff}}$ values from Gaia DR3 (Gaia Collaboration \etal 2023a). We applied upper limits for interstellar extinction toward the Galactic bulge, as calculated by Nataf \etal (2013).  In only a few cases, we utilized Galactic extinction maps by Gonzalez \etal (2012). \textsf{SED} modeling requires measurements of the star's bolometric flux in various bands. We reviewed publicly available photometric catalogs and their reference images for the Galactic bulge region. As a test case, we used the well-characterized planetary system OGLE-TR-10 (Udalski \etal 2002a, Konacki \etal 2003b, Melo \etal 2006), which is located in one of the analyzed fields, and its transit was detected in the \textsf{TLS} search. For this object, we prepared a set of modeling inputs in the same manner as for our candidates. Following \textsf{SED} modeling of OGLE-TR-10, we compared the fitted parameters to those reported in the literature. Ultimately, we decided to use measurements from the Dark Energy Camera Plane Survey 2  (DECaPS2, Saydjari \etal 2023), the Pan-STARRS1 Surveys DR2 (Chambers \etal 2016), and synthetic photometry from the Gaia DR3 low-resolution spectra (Gaia Collaboration \etal 2023b), as this set resulted in the best agreement between calculated parameters and those reported from high-resolution spectroscopy.
We assumed circular orbits for all of our objects since they are on short-period orbits. For stellar evolution models, we selected the default in \textsf{EXOFASTv2} - the MESA Isochrones and Stellar Tracks (\textsf{MIST}) model (Choi \etal 2016).
For each object, we performed a preliminary short modeling with parallel tempering. Subsequently, using the preliminary results, we recreated priors and conducted final modeling without parallel tempering and with 50~000 maximum steps.

\subsection{Final Selection}
To further reduce the likelihood of false positive signals in our catalog, we conducted additional tests on all objects for which we successfully created models. Our aim was to ensure that the transit signal in the light curve was not an artifact of a nearby eclipsing binary system. To achieve this, we cross-matched our candidates with the OGLE collection of eclipsing and ellipsoidal binary systems (Soszy\'nski \etal 2016). Within $1\arcm$ radius around  each candidate, we examined if there was a binary system with an orbital period or its alias consistent with that of our candidate. If such a match was found, we excluded the object from further analysis.
This was the case for 138  out of 1243 transits at this stage of the vetting process.
In the same manner as for the binary systems catalog, we cross-matched our final candidates with all the results from the \textsf{TLS} search. This additional step was designed to help eliminate false positive signals of instrumental origin and resulted in the exclusion of additional 110 detections.  Furthermore, we analyzed the histogram of periods detected by the \textsf{TLS} in each observed field and removed candidates with orbital periods close to the peaks of this histogram. An example section of the histogram from filed BLG501 is shown in Fig. 4. Those peaks vary between observation fields and can be associated with multiple astrophysical or instrumental signals, such as remaining aliases of the sidereal day, diffraction spikes from nearby bright stars, lunar cycle, or bad columns in the CCD detector.
We checked previously reported planetary-like transits from the third phase of the OGLE project (Udalski \etal 2002abc, 2008, Pont \etal 2008), and we did not find any matches with our detections beside the mentioned OGLE-TR-10.  OGLE-TR-10b is the only confirmed planet, which was regularly observed during OGLE-IV.
The absence of matches with previously unconfirmed candidates can be attributed to the more rigorous selection process.

To conclude the selection process, we chose the 99 most promising candidates for transiting planets, which will be presented in detail in Section 4. This final step considered various factors, including the number of observations, how well the created model fits the observational data, and whether the transit exhibits the V-shape, which could indicate a grazing binary system eclipse. 

\begin{figure}[h]
    \includegraphics[width=0.99\textwidth]{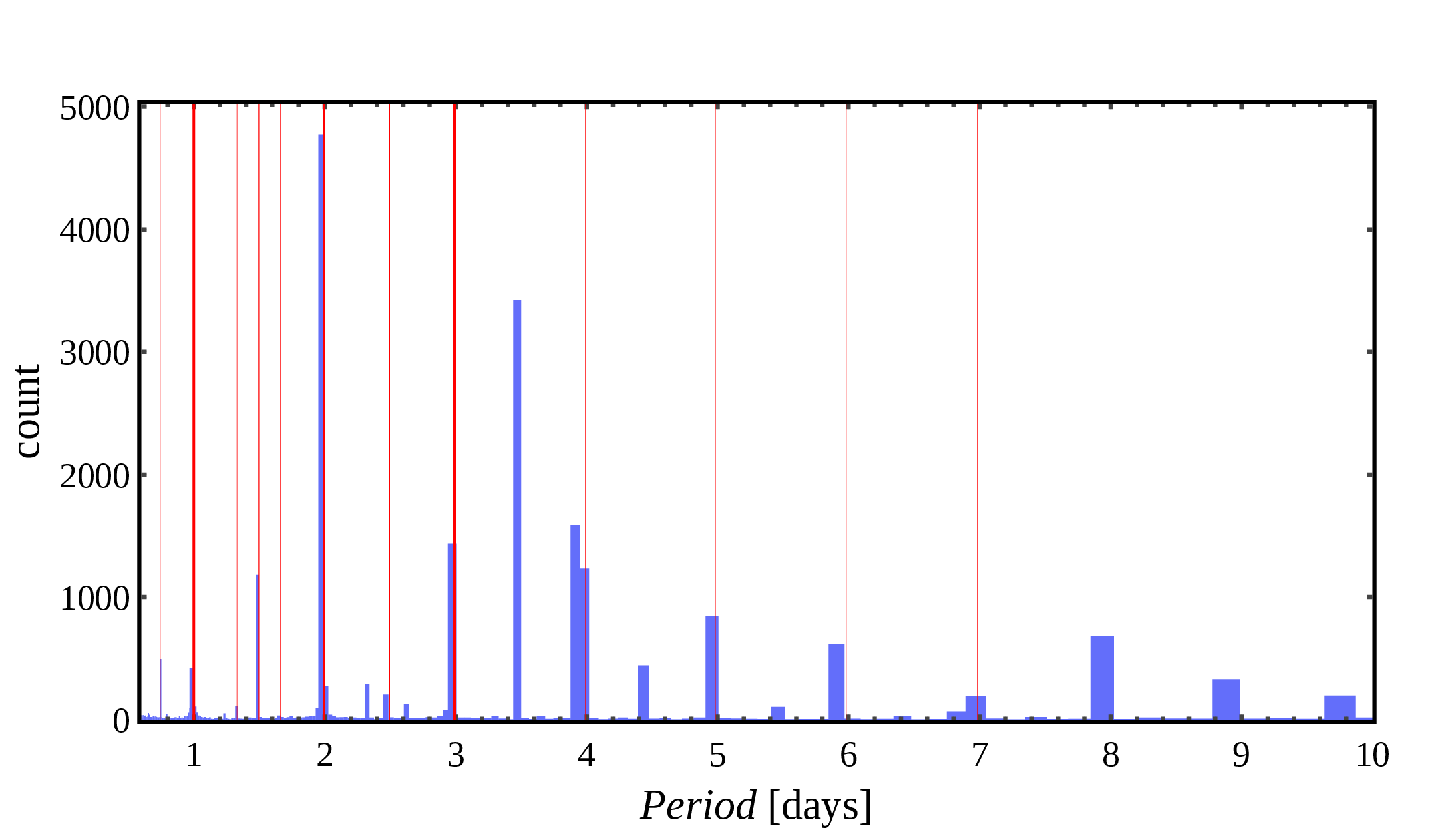}
    \FigCap{ Example section of the orbital periods distribution resulting from the \textsf{TLS} search in the field BLG501. Each bin represents 2000 values from the \textsf{TLS} period grid. The red fields indicate the excluded periods from Table 2.}{}
\end{figure}

\section{The Planetary Candidates}
Here, we present the selected planetary candidates. Table 3 displays their observational parameters, including celestial coordinates, orbital period of the companion ($P_{\rm{orb}}$), time of conjunction ($T_{\rm{c}}$), total transit duration ($T_{\rm{14}}$), fractional transit depth ($\delta$), brightness in $I$- and $V$-band, signal-to-noise ratio (SNR), and the probability that the transit is of planetary origin given the signal ($\rm{{Pr}(planet|signal)}$). Table 4 presents the selected physical parameters of the systems, including parameters of the host such as mass ($M_{*}$), radius ($R_{*}$), surface gravity  ($\rm{log}g$), effective temperature ($T_{\rm{eff}}$), metalicity  ($\rm{[Fe/H]}$), and distance  ($d$); parameters of the companion such as orbital radius  ($a$), inclination ($i$), and equilibrium temperature ($T_{\rm{eq}}$). For the detailed definitions of the parameters, we refer to Eastman \etal (2019). Figs. 7-13 show observational data of transits with fitted models. The identifiers of our planetary candidates follow the format OGLE-TR-NNNN, where NNNN represents a four-digit consecutive number starting from 1001 to distinguish from previous findings of the OGLE project (Udalski \etal 2002abc, 2008, Pont \etal 2008). 
The SNR is defined as:
\begin{equation}
    {\rm SNR} = \frac{\delta}{\sigma_o}n^{1/2} 
\end{equation}
where $\delta$ represents the transit depth, $\sigma$ is the standard deviation of the out-of-transit points, and $n$ is the number of in-transit points (Pont \etal 2006), Calculations of SNR values are based on \textsf{EXOFASTv2} models. All the selected candidates have SNR greater than 6, with median value $\rm{med}(\rm{SNR})= 20.46$.
The values of probability that the transit is of planetary origin $\rm{Pr(planet|signal)}$ were obtained with \textsf{VESPA}. As mentioned above, $\rm{Pr(planet|signal)}$  ranges between $0.8$ and $1.0$, with $\rm{Pr(planet|signal)}\geq 0.99$ for 63 objects .
The remaining parameters are median values resulting from the \textsf{MCMC} modeling (for detailed definitions of the parameters see Eastman \etal 2019).  
The majority of our candidates belong to the hot Jupiters family. The orbital periods range from $0.67$ d (slightly above the minimal period search by the \textsf{TLS} algorithm, $0.60$ d) to $63.22$ d. The radii of companions are typical for Jupiter-like gas giants, with a median value of $\rm{med}(R_{\rm{p}})=1.385~ R_{\rm{J}}$, with an exception of OGLE-TR-1003b. This candidate has the size corresponding to Neptune-like planets ($R_{\rm{p}}=0.529~R_{\rm{J}}$). The proximity to its host star places OGLE-TR-1003b in the parameter space of the so-called hot Neptune desert, which is an observed dearth of Neptune-size planets on orbits with periods shorter than~$\approx5 - 10$ d (Mazeh \etal 2016). In Fig. 5 we present  how the radii of candidates are distributed depending on the orbital period or equilibrium temperature, with the Neptune desert area marked. The distance distributions of the candidates span between 0.4 kpc and 5.5 kpc, which translates to $2.8 - 7.9$ kpc from the Galactic center, as shown in Fig. 6. Additionally, Fig. 6 illustrates the uniqueness of our findings in comparison to other known exoplanets in the Galaxy. To date, only observations from the Hubble Space Telescope have enabled the detection of sixteen transiting planetary candidates (two confirmed through RV measurements) close to the Galactic center (Sahu \etal 2006, 2008). The only  detection method that efficiently probes this region of the Galaxy is the microlensing method. Nevertheless, this method is
sensitive to planets on wider orbits compared to transit detections and, due to its transient nature, limits follow-up in-depth characterization of systems.

The full list of determined parameters of the systems with photometric data is available on the OGLE website \footnote[2]{ogledb.astrouw.edu.pl/\textasciitilde ogle/OCVS/}.

\begin{figure}[h!]
\includegraphics[width=0.99\textwidth]{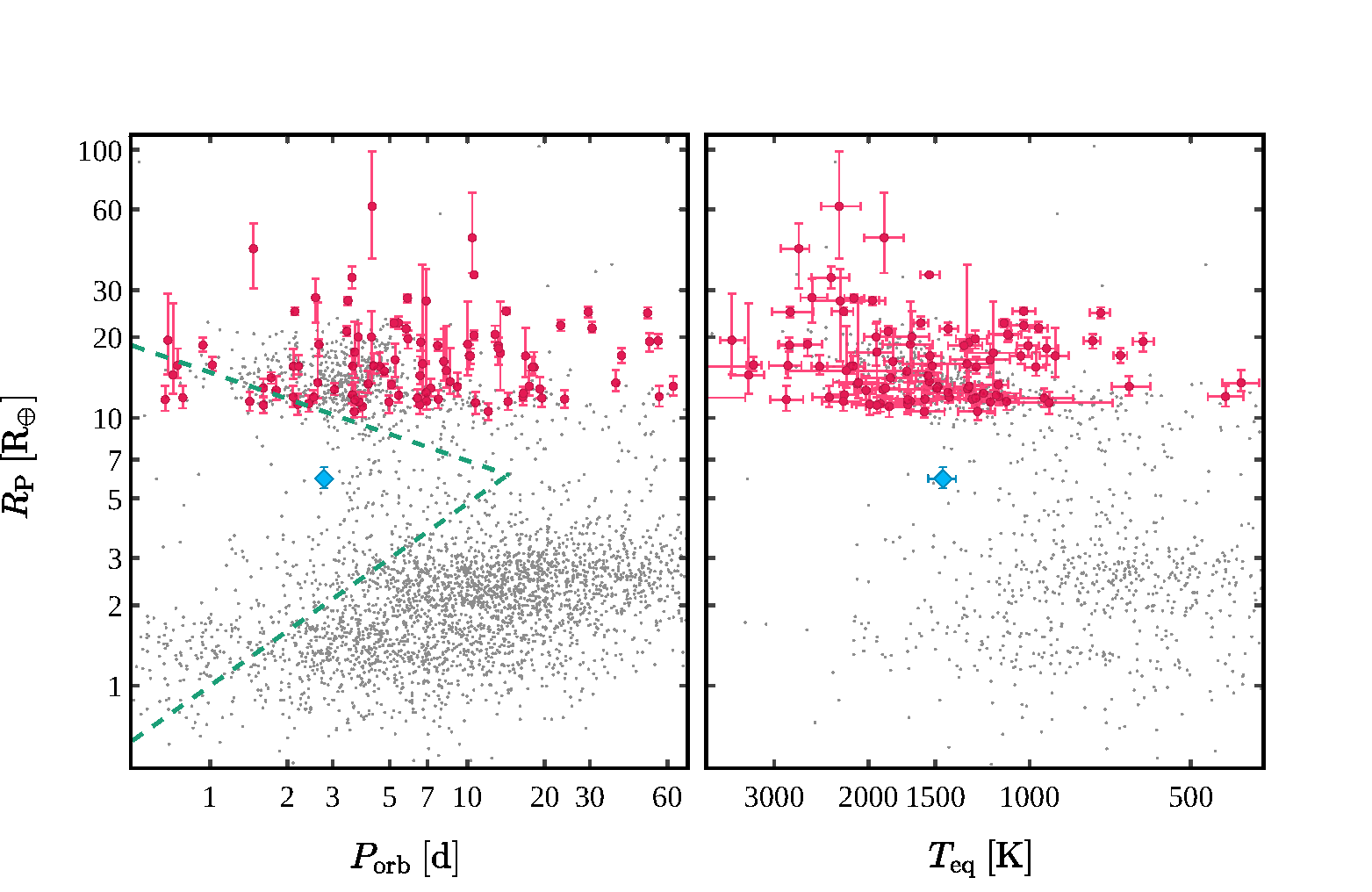}
\FigCap{Parameters of detected planetary candidates. The Jovian-size objects are
indicated by red points with error bars. Blue diamond mark the sub-Jovian object --
OGLE-TR-1003. Gray dots -- currently known confirmed planets (exoplanets.eu, as for September 2023).   The hot Neptune desert from Mazeh
\etal  (2016) shown with dashed lines in the left panel.  }
\end{figure}

\begin{figure}[h!]
\includegraphics[width=0.99\textwidth]{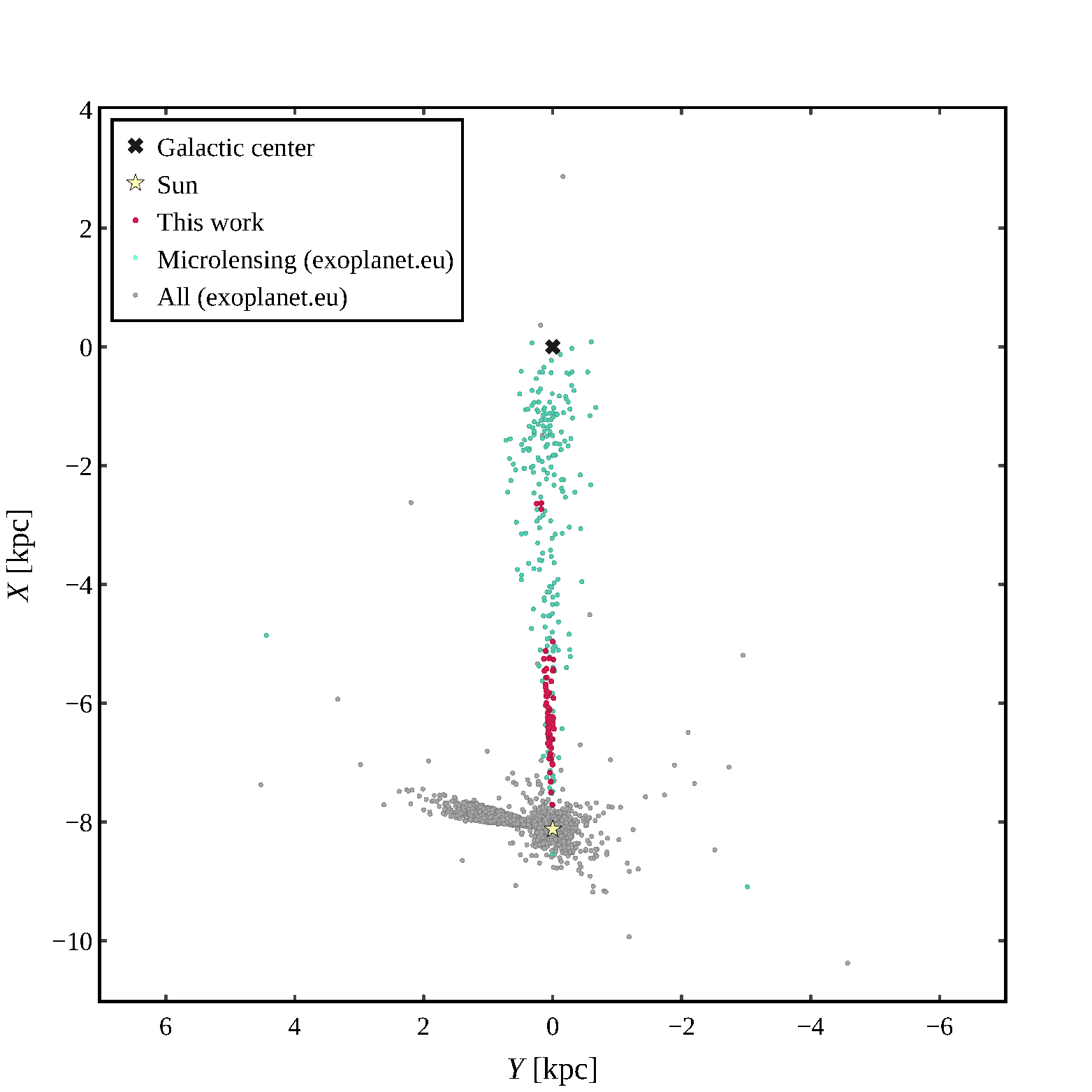}
\FigCap{Galactic distribution of our planetary candidates and known confirmed exoplanets (exoplanets.eu, as for September 2023).}
\end{figure}

\newpage

\section{Future Prospects}
We confirmed that observations of the Galactic bulge regions offer opportunities for planetary transit research. These regions are often overlooked by typical transit surveys due to the challenges associated with crowding. Nevertheless, they are and will continue to be monitored by surveys with a specific focus on detecting microlensing events. The large number of photometric measurements collected this way can be used for planetary transit studies.

One of the notable advantages of our catalog is the extended time span of available observations. For many of our candidates, we can expand the time coverage to an exceptional 23-year period by incorporating observations from the previous phase of the OGLE project (OGLE-III, 2001-2009). This long time span of regular observations has not been achieved by any other survey capable of detecting exoplanets. Consequently, our planetary candidates become valuable objects for studies of long-term orbital evolution (Hagey \etal 2022).  

Furthermore, the upcoming Nancy Grace Roman Space Telescope (Spergel \etal 2015, Montet \etal 2017) is expected to provide an additional 40~000 precise photometric and astrometric measurements of our targets. While our objects are brighter than the planned saturation limit of Roman ($14.8$ mag in filter $W149$, Penny \etal 2019), the expected relative precision will still be  of order of 1 part per thousand (Montet \etal 2017). Using noise estimation for a single epoch in the Roman $W149$ filter from Gould \etal (2015): 
\begin{equation}
  \sigma = 1.0 \times 10^{(2/15)(H-15)}~ \textrm{mmag}\quad \textrm{for} \quad 8\lesssim H \lesssim 15~\rm{mag}
\end{equation}
where $H$ stands for the near-infrared $H-$band, using values determined from  created \textsf{SED} models, we find that for our targets, $\sigma$ would be around $0.001$ mag.
\

As we mentioned in Subsection 3.2, the time-series observations alone are not enough to confirm the planetary nature of the objects. Given the relative faintness of our targets and the accessibility of telescopes, performing follow-up RV observations for all our candidates may not be achievable.  Nevertheless, this challenge exists since transit detections have increased in numbers, and various alternative validation methods have been developed, such as the probabilistic validation method used by us. Conducting additional photometric observations, including multi-band transit observations or high-resolution adaptive optics observations, will further help in establishing  the level of blended light. Additionally, obtaining low-resolution spectroscopic observations of the host stars can help narrow down their parameters and, indirectly, the parameters of the companions.

It is important to acknowledge that many of the selection steps for candidates presented in this work were subjective. Therefore, we advise exercising caution when using this catalog beyond a collection of individual cases for further analyses.

\section{Conclusions}
We conducted a planetary transit search using photometric observations collected during the fourth phase of the OGLE project.  We used observations of main sequence stars within four  fields covering a total area of $5.6 ~\rm{deg^2}$ in the direction of the Galactic bulge. To identify potential candidates, we employed a method that fits realistic transit shapes to data folded according to the grid of trial periods.  After the vetting process, which included the evaluation of the probability of false positive signals, we selected the most promising objects. Using all available data, sourced both from the OGLE project and the public domain, we developed comprehensive models of planetary systems. We presented selected 99 highly-probable planetary candidates. The majority of these candidates fall within the category of hot and warm Jupiters, with the exception of one object exhibiting a sub-Jovian size. Notably, this companion in OGLE-TR-1003 is located in the hot Neptune desert.
Our candidates are distributed across a wide range of distances, extending up to 5.5 kpc, dovetailing with planets discovered with the microlensing method.  This synergy offers an extraordinary opportunity for the study of planetary systems, encompassing those with the shortest to the longest orbital periods as a function of Galactic distance. Furthermore, finding transiting planets in fields monitored by microlensing surveys is extremely beneficial in terms of time coverage of observations.  In the case of our candidates, the OGLE database can provide regular and consistent observations since  2001, with the potential for further extensions as the project continues. Additionally, the upcoming Nancy Grace Roman Space Telescope mission is expected to provide additional observations of our targets with improved precision.

\Acknow{
We thank all the OGLE observers for their contribution to the collection of the photometric data over the decades. 

This work has been funded by the National Science Centre, Poland, grant
no.~2022/45/B/ST9/00243. For the purpose of Open Access, the author has
applied a CC-BY public copyright license to any Author Accepted Manuscript
(AAM) version arising from this submission. 

We used data from the European Space Agency (ESA) mission Gaia, processed by the Gaia Data Processing and Analysis Consortium (DPAC). Funding for the DPAC has been provided by national institutions, in particular the institutions participating in the Gaia Multilateral Agreement.

This project used public archival data from the Dark Energy Survey (DES). Funding for the DES Projects has been provided by the U.S. Department of Energy, the U.S. National Science Foundation, the Ministry of Science and Education of Spain, the Science and Technology Facilities Council of the United Kingdom, the Higher Education Funding Council for England, the National Center for Supercomputing Applications at the University of Illinois at Urbana-Champaign, the Kavli Institute of Cosmological Physics at the University of Chicago, the Center for Cosmology and Astro-Particle Physics at the Ohio State University, the Mitchell Institute for Fundamental Physics and Astronomy at Texas A\&M University, Financiadora de Estudos e Projetos, Funda{\c c}{\~a}o Carlos Chagas Filho de Amparo {\`a} Pesquisa do Estado do Rio de Janeiro, Conselho Nacional de Desenvolvimento Cient{\'i}fico e Tecnol{\'o}gico and the Minist{\'e}rio da Ci{\^e}ncia, Tecnologia e Inova{\c c}{\~a}o, the Deutsche Forschungsgemeinschaft, and the Collaborating Institutions in the Dark Energy Survey.
The Collaborating Institutions are Argonne National Laboratory, the University of California at Santa Cruz, the University of Cambridge, Centro de Investigaciones Energ{\'e}ticas, Medioambientales y Tecnol{\'o}gicas-Madrid, the University of Chicago, University College London, the DES-Brazil Consortium, the University of Edinburgh, the Eidgen{\"o}ssische Technische Hochschule (ETH) Z{\"u}rich,  Fermi National Accelerator Laboratory, the University of Illinois at Urbana-Champaign, the Institut de Ci{\`e}ncies de l'Espai (IEEC/CSIC), the Institut de F{\'i}sica d'Altes Energies, Lawrence Berkeley National Laboratory, the Ludwig-Maximilians Universit{\"a}t M{\"u}nchen and the associated Excellence Cluster Universe, the University of Michigan, the National Optical Astronomy Observatory, the University of Nottingham, The Ohio State University, the OzDES Membership Consortium, the University of Pennsylvania, the University of Portsmouth, SLAC National Accelerator Laboratory, Stanford University, the University of Sussex, and Texas A\&M University.
Based in part on observations at Cerro Tololo Inter-American Observatory, National Optical Astronomy Observatory, which is operated by the Association of Universities for Research in Astronomy (AURA) under a cooperative agreement with the National Science Foundation.

The Pan-STARRS1 Surveys (PS1) and the PS1 public science archive have been made possible through contributions by the Institute for Astronomy, the University of Hawaii, the Pan-STARRS Project Office, the Max-Planck Society and its participating institutes, the Max Planck Institute for Astronomy, Heidelberg and the Max Planck Institute for Extraterrestrial Physics, Garching, The Johns Hopkins University, Durham University, the University of Edinburgh, the Queen's University Belfast, the Harvard-Smithsonian Center for Astrophysics, the Las Cumbres Observatory Global Telescope Network Incorporated, the National Central University of Taiwan, the Space Telescope Science Institute, the National Aeronautics and Space Administration under Grant No. NNX08AR22G issued through the Planetary Science Division of the NASA Science Mission Directorate, the National Science Foundation Grant No. AST-1238877, the University of Maryland, Eotvos Lorand University (ELTE), the Los Alamos National Laboratory, and the Gordon and Betty Moore Foundation.}

\newpage
\begin{figure}[htb]
\includegraphics[width=0.98\textwidth]{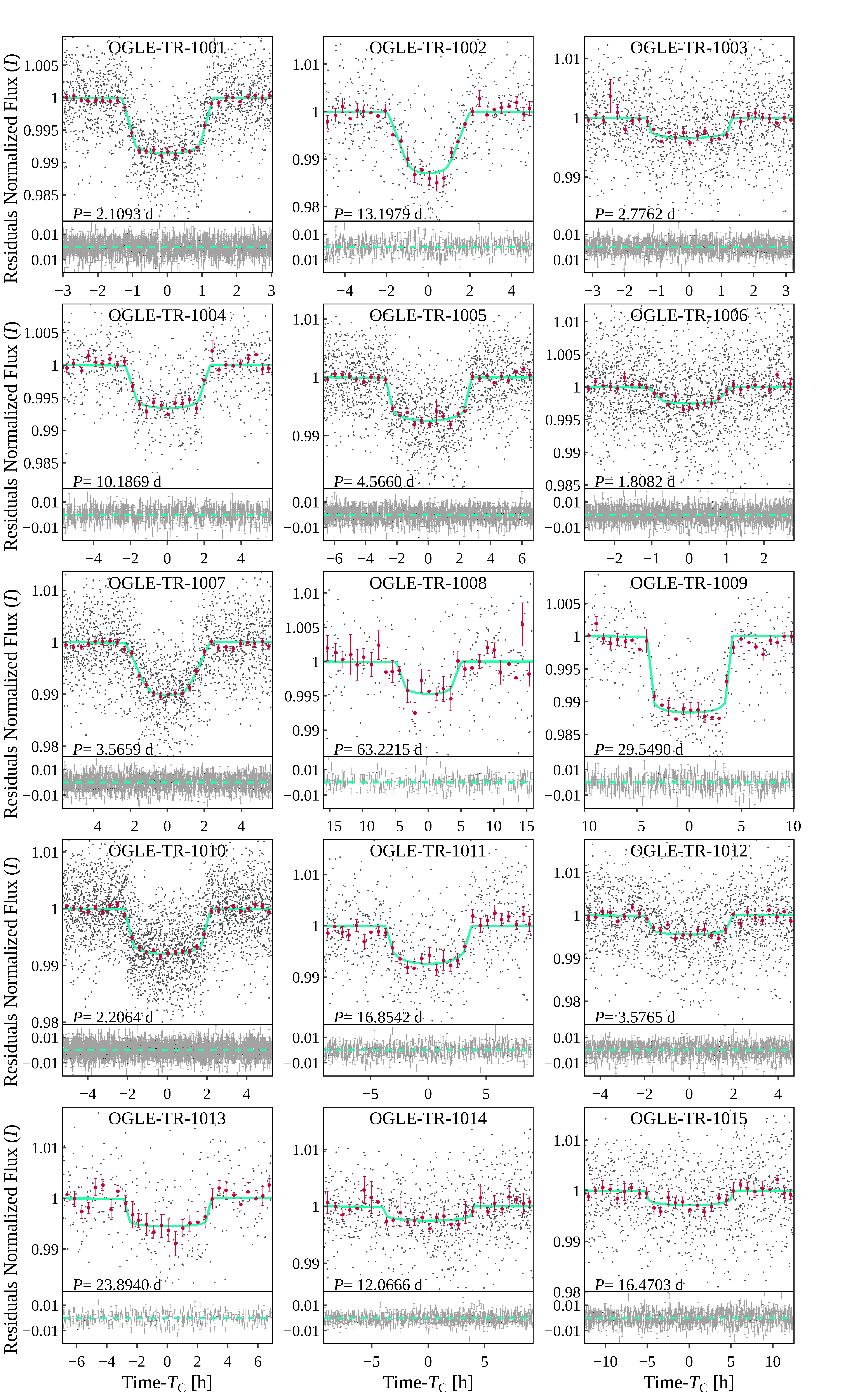}
\FigCap{Light curves of planetary transit candidates: OGLE-TR-1001--OGLE-TR-1015. Gray points -- folded observational data, red points -- binned observational data, green line -- fitted \textsf{MCMC} model with minima $\chi^2$.}
\end{figure}

\newpage
\begin{figure}[htb]
\includegraphics[width=0.98\textwidth]{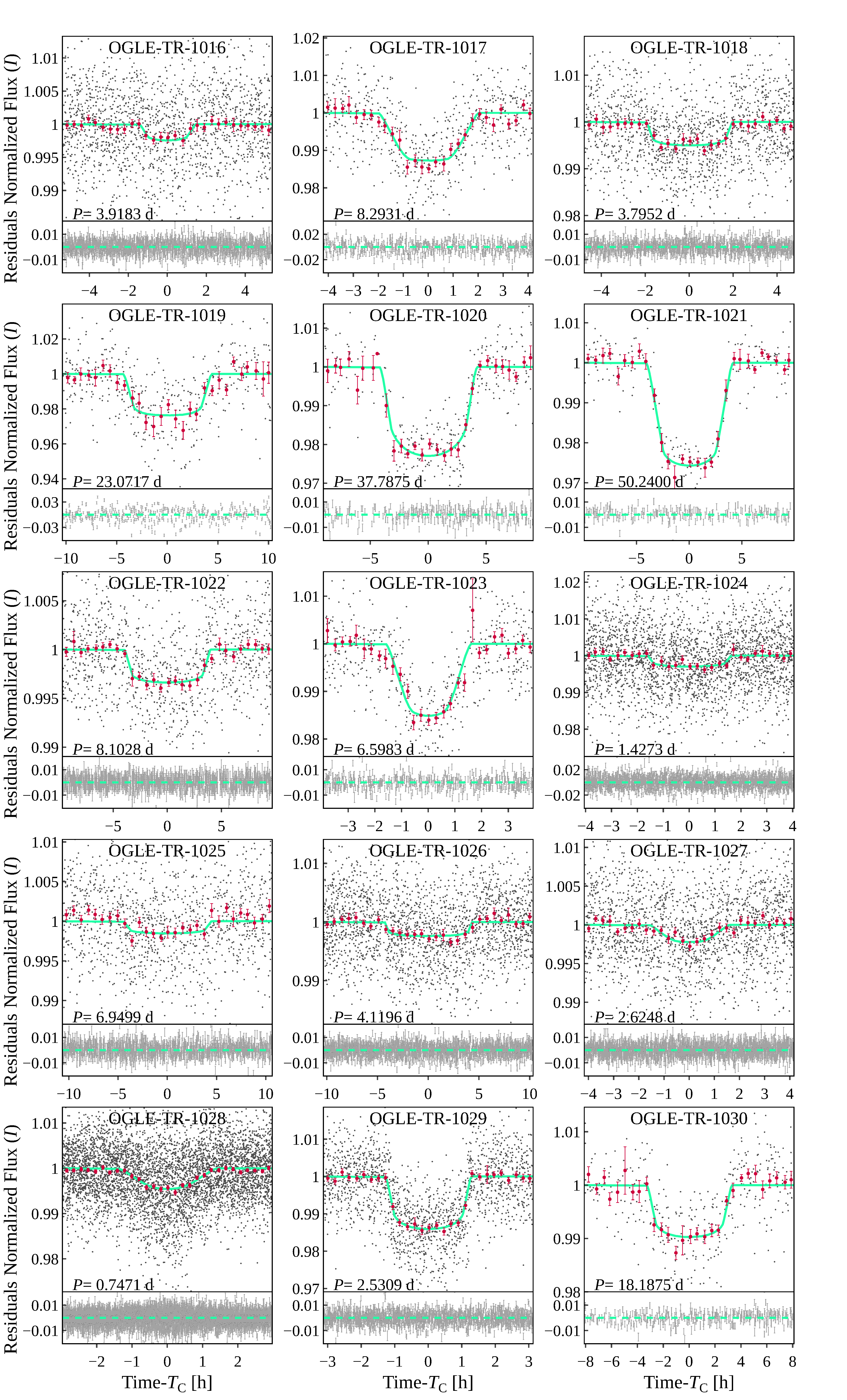}
\FigCap{Same as Fig. 7, planetary transit candidates: OGLE-TR-1016--OGLE-TR-1030.}
\end{figure}

\newpage
\begin{figure}[htb]
\includegraphics[width=0.98\textwidth]{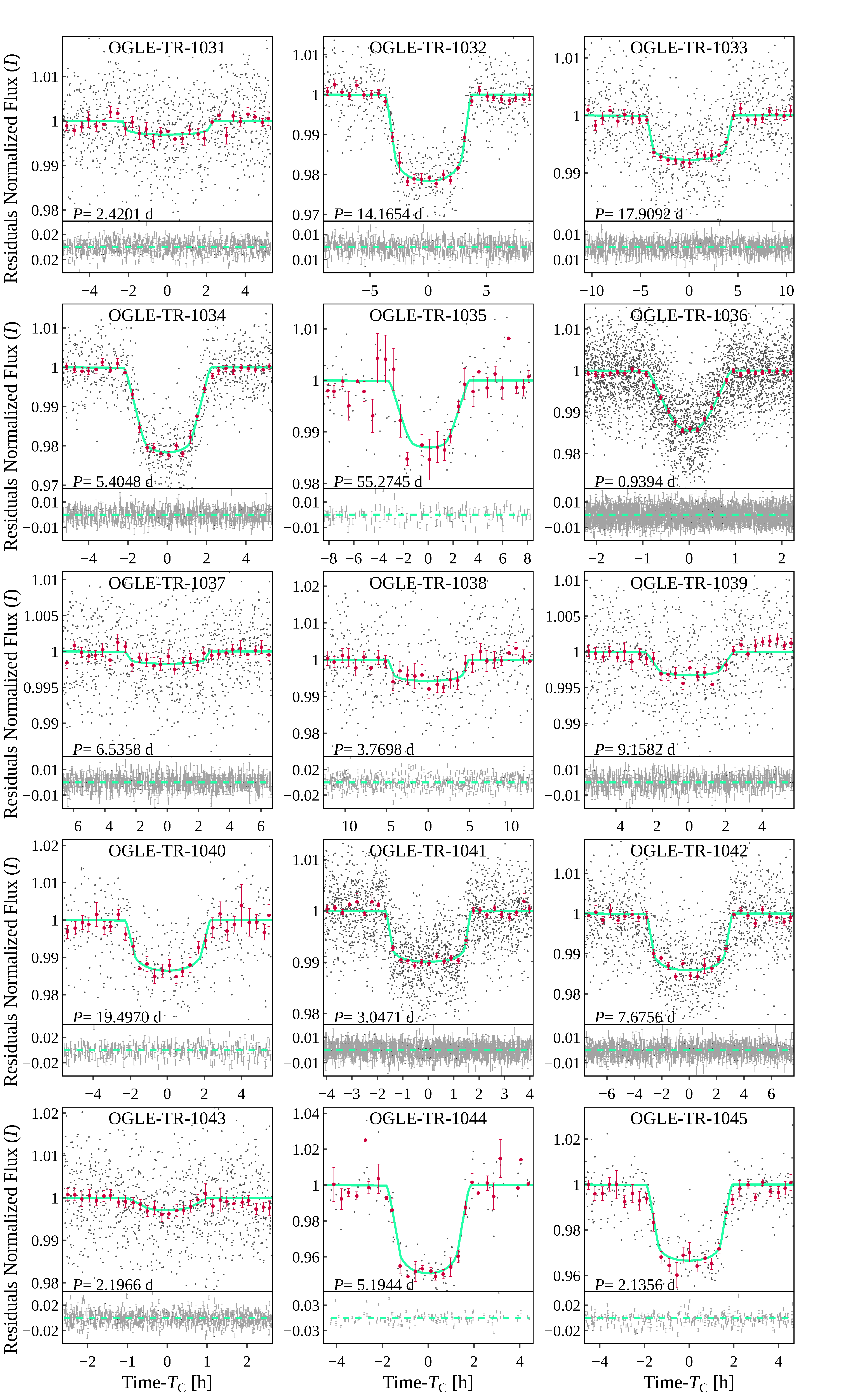}
\FigCap{Same as Fig. 7, planetary transit candidates: OGLE-TR-1031--OGLE-TR-1045.}
\end{figure}

\newpage
\begin{figure}[htb]
\includegraphics[width=0.98\textwidth]{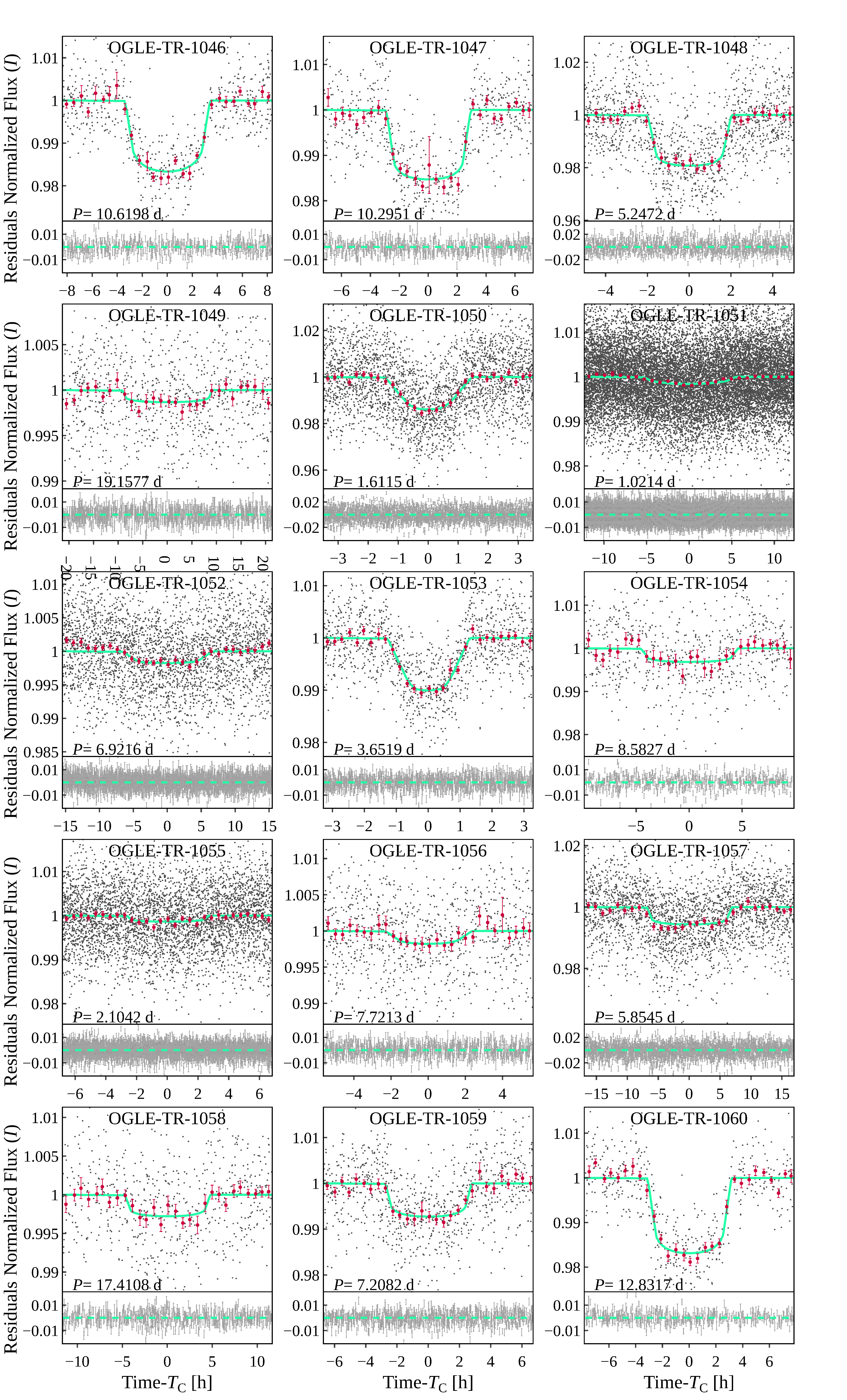}
\FigCap{Same as Fig. 7, planetary transit candidates: OGLE-TR-1046--OGLE-TR-1060.}
\end{figure}

\newpage
\begin{figure}[htb]
\includegraphics[width=0.98\textwidth]{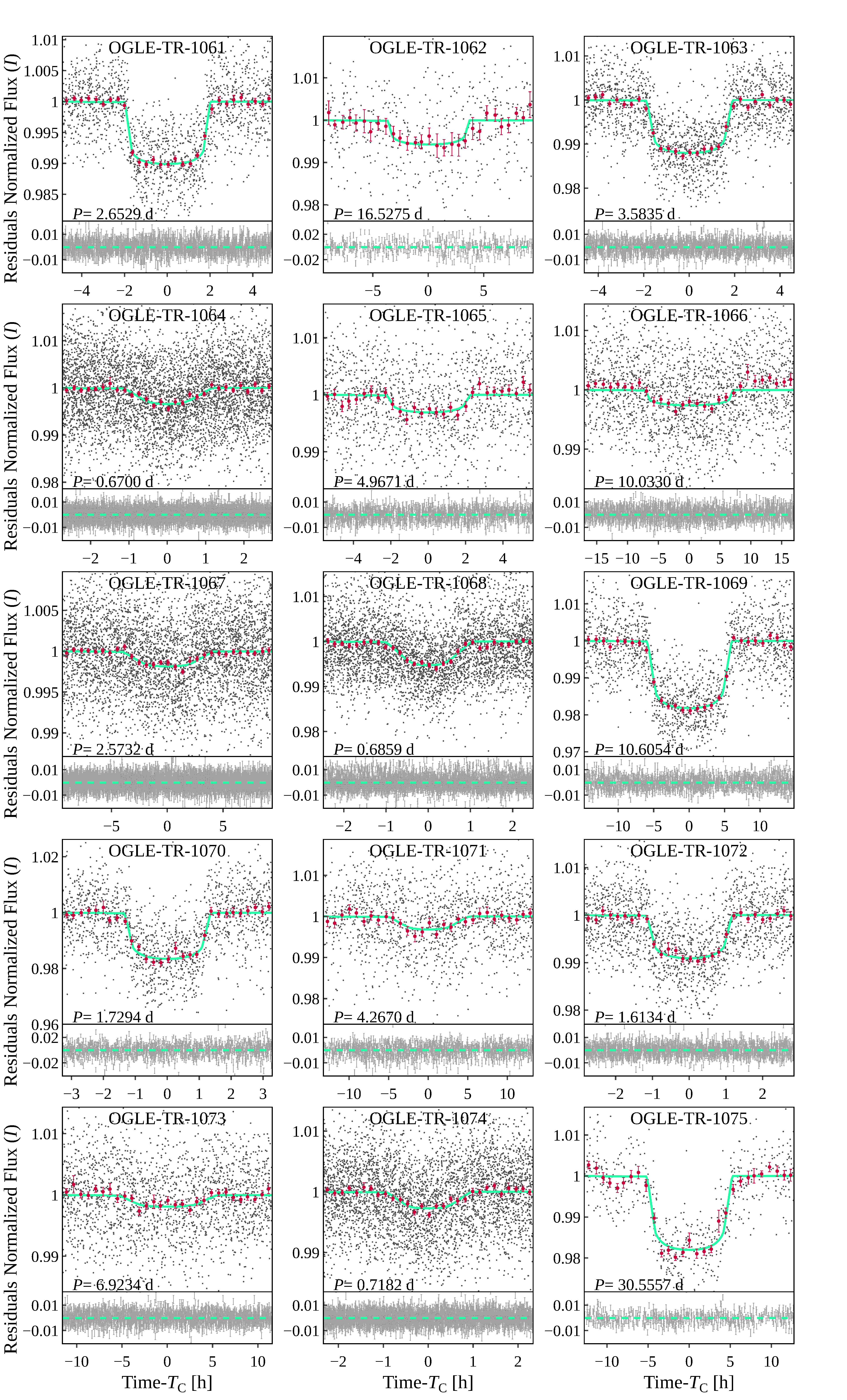}
\FigCap{Same as Fig. 7, planetary transit candidates: OGLE-TR-1061--OGLE-TR-1075.}
\end{figure}

\newpage
\begin{figure}[htb]
\includegraphics[width=0.98\textwidth]{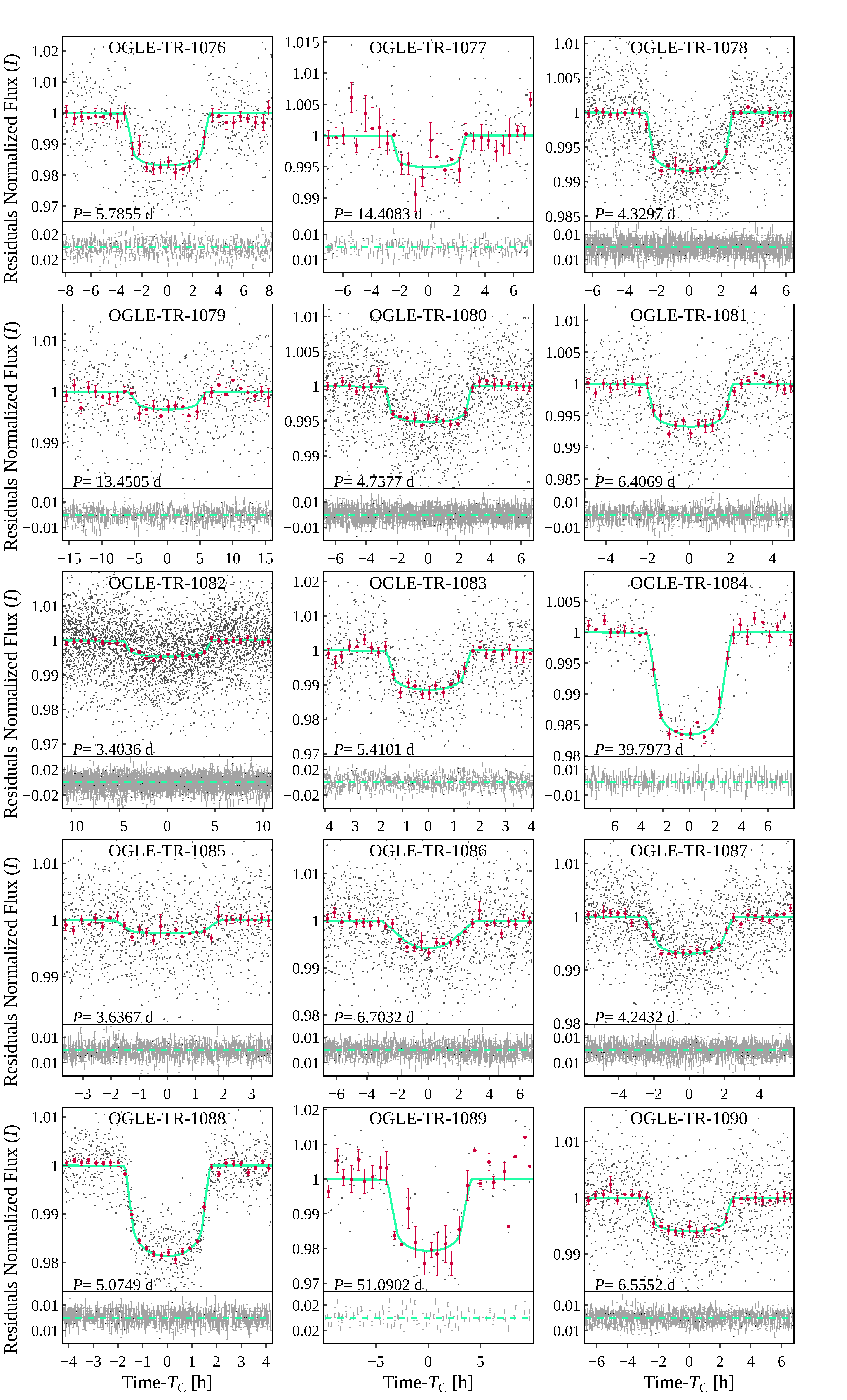}
\FigCap{Same as Fig. 7, planetary transit candidates: OGLE-TR-1076--OGLE-TR-1090.}
\end{figure}

\newpage
\begin{figure}[ht]
\begin{center}
\includegraphics[width=0.98\textwidth]{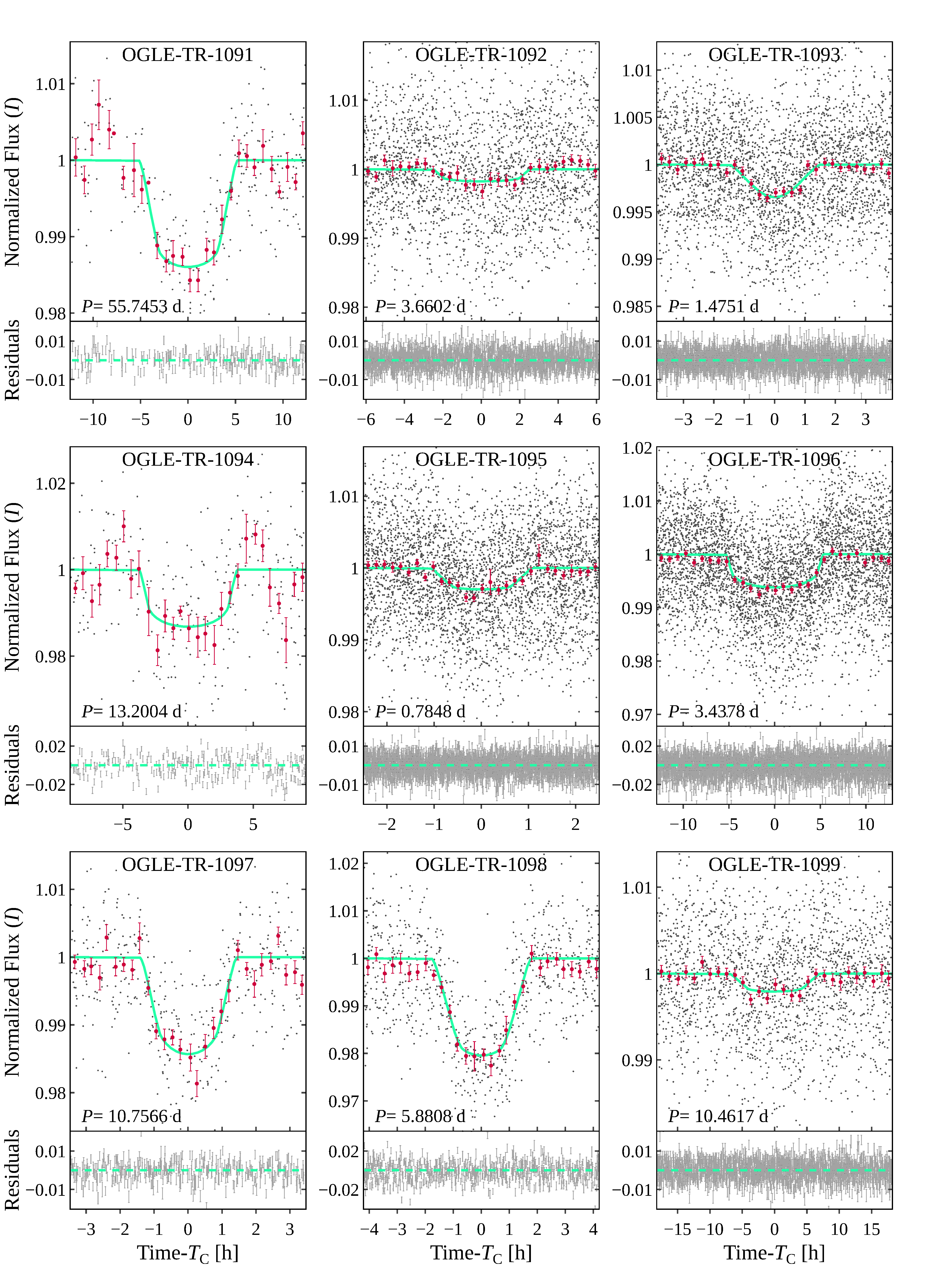}
\end{center}
\FigCap{Same as Fig. 7, planetary transit candidates: OGLE-TR-1091--OGLE-TR-1099.}
\end{figure}

\newpage
\begin{landscape}
{
\TabCapp{12pt}{Observational parameters of systems with planetary candidates.}
\footnotesize
\tablefirsthead{
\hline\noalign{\smallskip}
Name & RA & DEC & $P_{\rm{orb}}$ & $T_{\rm{c}}$& $T_{\rm{14}}$ & $\delta$ & $I$ &$V$ & $\rm{SNR}$ & $\rm{{Pr}(planet | signal)}$ \\ &(J2000) &(J2000) & [d] & [$\rm{BJD_{TDB}} - 2450000$] & [d] & & [mag] & [mag]& & \cr\noalign{\smallskip}
\hline\noalign{\smallskip} }
\tablehead{
\multicolumn{11}{c}{\normalsize{  {\spaceskip 2pt plus 1pt minus 1pt T a b l e 3}}
}\\ \noalign{\medskip} \multicolumn{11}{c}{ \normalsize{ {Continued }}} \\\noalign{\bigskip}
\hline\noalign{\smallskip}
Name & RA & DEC & $P_{\rm{orb}}$ & $T_{\rm{c}}$& $T_{\rm{14}}$ & $\delta$ & $I$ &$V$ & $\rm{SNR}$ & $\rm{{Pr}(planet | signal)}$ \\ &(J2000) &(J2000) & [d] & [$\rm{BJD_{TDB}} - 2450000$] & [d] & & [mag] & [mag]& & \cr\noalign{\smallskip}
\hline \noalign{\smallskip}
}
\tabletail{
\noalign{\smallskip}\hline
\multicolumn{11}{c}{\textit{continued on next page}}\\
}
\tablelasttail{\hline}
\begin{supertabular}{ccccccccccc}
OGLE-TR-1001 & $17:49:43.17$ & $-29:23:41.9$ & $2.1093174$ & $5378.18379$ & $0.1043$ &  $0.00823$ &    13.923 &    14.815 & 58.05 & 1.00 \\
OGLE-TR-1002 & $17:49:51.28$ & $-29:18:01.1$ &        $13.197893$ &             $5379.7864$ &        $0.1740$ &   $0.0142$ &    15.419 &    17.021 & 32.86 & 0.93 \\
OGLE-TR-1003 & $17:50:22.77$ & $-29:23:19.4$ &        $2.7762253$ &             $5379.3403$ &        $0.1123$ &  $0.00285$ &    15.382 &    16.872 & 12.60 &              1.00 \\
OGLE-TR-1004 & $17:50:23.35$ & $-30:15:01.4$ &        $10.186892$ &             $5379.8420$ &        $0.1964$ &  $0.00635$ &    14.168 &    15.348 & 29.93 &              1.00 \\
OGLE-TR-1005 & $17:50:28.70$ & $-29:54:23.9$ &        $4.5659828$ &             $5380.5474$ &        $0.2317$ &  $0.00655$ &    14.824 &    15.931 & 35.43 &              1.00 \\
OGLE-TR-1006 & $17:50:38.40$ & $-29:22:06.8$ &        $1.8082001$ &             $5377.0093$ &        $0.0970$ &  $0.00287$ &    14.902 &    16.640 & 16.33 &              0.99 \\
OGLE-TR-1007 & $17:50:47.56$ & $-30:11:59.7$ &        $3.5659243$ &             $5379.2049$ &        $0.1960$ &  $0.01082$ &    12.855 &    13.626 & 56.25 &              1.00 \\
OGLE-TR-1008 & $17:50:48.36$ & $-29:40:49.1$ &          $63.2215$ &              $5416.535$ &         $0.552$ &  $0.00331$ &    15.056 &    16.583 &  6.39 &              1.00 \\
OGLE-TR-1009 & $17:50:56.70$ & $-29:40:49.1$ &        $29.549047$ &             $5398.1766$ &        $0.3472$ &  $0.01088$ &    14.155 &    15.146 & 36.99 &              1.00 \\
OGLE-TR-1010 & $17:51:30.88$ & $-29:55:45.2$ &        $2.2063895$ &             $5377.4247$ &        $0.1826$ &  $0.00699$ &    14.519 &    15.718 & 51.76 &              1.00 \\
OGLE-TR-1011 & $17:51:53.22$ & $-30:01:15.7$ &        $16.854160$ &             $5389.1759$ &         $0.313$ &  $0.00654$ &    15.463 &    17.080 & 15.91 &              0.99 \\
OGLE-TR-1012 & $17:52:14.08$ & $-29:14:41.8$ &         $3.576539$ &             $5379.8131$ &        $0.1629$ &  $0.00396$ &    15.365 &    16.549 & 14.31 &              1.00 \\
OGLE-TR-1013 & $17:52:27.45$ & $-29:53:02.5$ &         $23.89398$ &              $5390.170$ &         $0.240$ &  $0.00538$ &    15.031 &    16.035 &  9.03 &              0.99 \\
OGLE-TR-1014 & $17:52:29.08$ & $-29:25:30.7$ &         $12.06657$ &              $5379.955$ &         $0.321$ &  $0.00254$ &    14.280 &    15.423 &  7.05 &              0.98 \\
OGLE-TR-1015 & $17:52:32.93$ & $-30:03:59.9$ &         $16.47033$ &              $5377.320$ &         $0.433$ &  $0.00241$ &    15.432 &    16.491 &  7.33 &              1.00 \\
OGLE-TR-1016 & $17:52:47.03$ & $-29:53:32.2$ &         $3.918334$ &              $5377.401$ &         $0.186$ &  $0.00184$ &    15.122 &    16.191 &  8.98 &              0.88 \\
OGLE-TR-1017 & $17:53:03.85$ & $-30:06:36.0$ &         $8.293105$ &             $5378.3103$ &        $0.1452$ &   $0.0121$ &    15.172 &    16.462 & 23.69 &              1.00 \\
OGLE-TR-1018 & $17:54:28.67$ & $-30:04:39.9$ &         $3.795212$ &             $5376.7046$ &        $0.1650$ &  $0.00446$ &    15.380 &    16.568 & 15.26 &              1.00 \\
OGLE-TR-1019 & $17:54:35.45$ & $-27:43:17.3$ &         $23.07172$ &             $5377.6698$ &         $0.358$ &   $0.0215$ &    15.306 &    17.194 & 17.85 &              0.92 \\
OGLE-TR-1020 & $17:55:01.28$ & $-29:27:10.0$ &         $37.78747$ &             $5388.5955$ &         $0.313$ &   $0.0202$ &    15.348 &    17.030 & 38.64 &              1.00 \\
OGLE-TR-1021 & $17:55:17.82$ & $-27:37:51.7$ &         $50.24001$ &             $5394.4031$ &        $0.3439$ &  $0.02392$ &    14.680 &    15.917 & 36.94 &              0.98 \\
OGLE-TR-1022 & $17:55:17.87$ & $-27:52:33.5$ &         $8.102826$ &             $5383.7016$ &         $0.335$ &  $0.00336$ &    14.119 &    15.414 & 21.31 &              0.99 \\
OGLE-TR-1023 & $17:55:18.63$ & $-28:22:08.5$ &         $6.598261$ &             $5378.3482$ &        $0.1358$ &   $0.0161$ &    15.220 &    16.799 & 36.96 &              1.00 \\
OGLE-TR-1024 & $17:55:28.73$ & $-27:26:23.4$ &        $1.4272776$ &             $5377.3506$ &        $0.1400$ &  $0.00269$ &    15.145 &    16.774 &  9.66 &              0.94 \\
OGLE-TR-1025 & $17:55:41.38$ & $-27:29:06.8$ &         $6.949874$ &              $5377.984$ &         $0.368$ &  $0.00138$ &    14.727 &    16.192 &  7.64 &              0.84 \\ 
OGLE-TR-1026 &  $17:55:43.01$  & $ -27:55:57.9 $ &         $4.119552$ &              $5376.948$ &         $0.357$ &  $0.00208$ &    14.528 &    15.635 & 10.82 &              0.97 \\
OGLE-TR-1027 & $17:55:49.63$ & $-29:11:39.3$ &         $2.624793$ &              $5261.407$ &         $0.144$ &  $0.00203$ &    14.785 &    16.199 & 12.84 &              0.82 \\
OGLE-TR-1028 & $17:55:56.66$ & $-29:17:03.8$ &       $0.74714108$ &             $5376.7574$ &        $0.1027$ &  $0.00489$ &    13.753 &    15.111 & 43.03 &              0.90 \\
OGLE-TR-1029 & $17:56:00.49$ & $-28:49:24.8$ &        $2.5308880$ &             $5261.7539$ &        $0.1081$ &  $0.01217$ &    15.194 &    16.400 & 44.07 &              1.00 \\
OGLE-TR-1030 & $17:56:05.22$ & $-27:56:09.0$ &        $18.187463$ &             $5388.3494$ &         $0.280$ &  $0.00841$ &    15.276 &    16.945 & 18.11 &              0.99 \\
OGLE-TR-1031 & $17:56:11.67$ & $-27:40:02.7$ &         $2.420147$ &              $5376.749$ &         $0.186$ &  $0.00280$ &    14.337 &    15.605 &  7.38 &              1.00 \\
OGLE-TR-1032 & $17:56:16.96$ & $-28:24:39.9$ &        $14.165438$ &             $5383.9723$ &        $0.3117$ &  $0.01825$ &    15.242 &    16.913 & 50.46 &              1.00 \\
OGLE-TR-1033 & $17:56:22.42$ & $-29:10:17.1$ &        $17.909180$ &             $5381.5566$ &        $0.3724$ &  $0.00680$ &    15.296 &    16.933 & 25.54 &              1.00 \\
OGLE-TR-1034 & $17:56:39.10$ & $-27:36:45.3$ &        $5.4047713$ &             $5377.5239$ &        $0.1843$ &  $0.02099$ &    15.194 &    16.986 & 62.69 &              0.98 \\
OGLE-TR-1035 & $17:56:42.84$ & $-28:19:34.9$ &         $55.27448$ &              $5379.163$ &         $0.292$ &   $0.0124$ &    13.646 &    14.566 & 15.59 &              0.98 \\
OGLE-TR-1036 & $17:56:47.29$ & $-28:37:15.7$ &       $0.93942100$ &            $5261.03860$ &        $0.0782$ &  $0.01648$ &    14.029 &    14.911 & 96.27 &              0.92 \\
OGLE-TR-1037 & $17:57:00.17$ & $-28:20:07.3$ &         $6.535826$ &              $5381.591$ &         $0.232$ &  $0.00162$ &    14.706 &    15.982 &  7.21 &              0.86 \\
OGLE-TR-1038 & $17:57:12.92$ & $-27:29:09.3$ &         $3.769846$ &              $5379.148$ &         $0.436$ &  $0.00477$ &    13.630 &    14.904 &  6.95 &              1.00 \\
OGLE-TR-1039 & $17:57:19.05$ & $-29:49:41.8$ &         $9.158188$ &             $5380.9316$ &        $0.1984$ &  $0.00325$ &    14.900 &    16.115 & 13.59 &              0.94 \\
OGLE-TR-1040 & $17:57:21.04$ & $-28:49:13.1$ &        $19.496967$ &             $5277.9227$ &        $0.1955$ &  $0.01203$ &    15.394 &    16.719 & 18.42 &              1.00 \\
OGLE-TR-1041 & $17:57:27.61$ & $-29:30:00.4$ &        $3.0470709$ &             $5377.2418$ &        $0.1426$ &  $0.00881$ &    14.996 &    16.394 & 45.77 &              1.00 \\
OGLE-TR-1042 & $17:57:29.88$ & $-28:42:17.1$ &         $7.675599$ &             $5267.8518$ &        $0.2646$ &  $0.01221$ &    15.492 &    17.101 & 39.99 &              1.00 \\
OGLE-TR-1043 & $17:57:44.37$ & $-27:58:28.5$ &         $2.196576$ &               $5378.07$ &         $0.091$ &   $0.0031$ &    15.359 &    17.073 &  7.24 &              0.98 \\
OGLE-TR-1044 & $17:57:54.98$ & $-27:48:40.8$ &        $5.1944381$ &             $5380.3002$ &        $0.1583$ &   $0.0417$ &    13.836 &    14.852 & 21.54 &              1.00 \\
OGLE-TR-1045 & $17:57:55.52$ & $-29:13:04.2$ &        $2.1355658$ &             $5262.4665$ &        $0.1623$ &   $0.0302$ &    14.730 &    15.750 & 26.05 &              1.00 \\
OGLE-TR-1046 & $17:57:56.84$ & $-27:36:50.5$ &        $10.619790$ &             $5381.1543$ &        $0.2896$ &  $0.01414$ &    15.245 &    16.751 & 34.92 &              0.96 \\
OGLE-TR-1047 & $17:58:06.08$ & $-27:46:06.7$ &        $10.295061$ &             $5378.3788$ &        $0.2506$ &  $0.01349$ &    15.476 &    16.759 & 33.56 &              1.00 \\
OGLE-TR-1048 & $17:58:07.33$ & $-29:27:32.2$ &        $5.2472019$ &             $5378.6899$ &        $0.1735$ &  $0.01730$ &    15.431 &    16.736 & 32.92 &              1.00 \\
OGLE-TR-1049 & $17:58:17.11$ & $-27:34:26.1$ &         $19.15773$ &              $5384.334$ &         $0.737$ &  $0.00122$ &    14.629 &    16.007 &  6.33 &              1.00 \\
OGLE-TR-1050 & $17:58:27.67$ & $-29:49:32.0$ &        $1.6114682$ &             $5377.8440$ &        $0.1208$ &   $0.0152$ &    15.269 &    16.204 & 41.03 &              0.97 \\
OGLE-TR-1051 & $17:58:31.55$ & $-29:16:55.4$ &        $1.0213727$ &              $5376.673$ &         $0.465$ & $0.001312$ &    15.324 &    17.065 & 17.85 &              0.97 \\
OGLE-TR-1052 & $17:58:33.17$ & $-29:23:01.1$ &         $6.921614$ &              $5377.348$ &         $0.534$ &  $0.00169$ &    14.876 &    16.149 & 12.62 &              0.96 \\
OGLE-TR-1053 & $17:58:34.08$ & $-28:48:24.9$ &        $3.6519106$ &             $5273.7362$ &        $0.1135$ &   $0.0109$ &    14.915 &    15.829 & 47.88 &              1.00 \\
OGLE-TR-1054 & $17:58:39.75$ & $-27:22:32.0$ &          $8.58268$ &              $5378.537$ &         $0.342$ &  $0.00272$ &    15.066 &    16.220 &  6.25 &              0.98 \\
OGLE-TR-1055 & $17:58:41.82$ & $-29:11:27.7$ &         $2.104221$ &              $5503.963$ &         $0.236$ &  $0.00135$ &    14.919 &    16.800 &  7.99 &              0.95 \\
OGLE-TR-1056 & $17:58:55.74$ & $-28:53:03.8$ &           $7.7213$ &                $5263.3$ &         $0.195$ &  $0.00181$ &    14.913 &    16.686 &  7.03 &              0.94 \\
OGLE-TR-1057 & $17:59:05.12$ & $-28:29:51.9$ &         $5.854539$ &             $5378.9814$ &         $0.586$ &  $0.00491$ &    15.245 &    17.114 & 18.57 &              1.00 \\
OGLE-TR-1058 & $17:59:14.69$ & $-27:44:55.6$ &         $17.41082$ &              $5382.036$ &         $0.403$ &  $0.00217$ &    14.806 &    16.222 &  8.02 &              0.94 \\
OGLE-TR-1059 & $17:59:24.87$ & $-28:01:37.4$ &         $7.208224$ &             $5377.8502$ &        $0.2321$ &  $0.00646$ &    15.412 &    16.726 & 19.09 &              0.97 \\
OGLE-TR-1060 & $17:59:27.20$ & $-28:00:26.8$ &        $12.831750$ &             $5386.0147$ &        $0.2709$ &  $0.01508$ &    15.307 &    16.677 & 34.86 &              1.00 \\
OGLE-TR-1061 & $17:59:30.54$ & $-27:23:20.3$ &        $2.6528610$ &             $5376.6303$ &        $0.1696$ &  $0.00915$ &    14.119 &    14.767 & 55.65 &              1.00 \\
OGLE-TR-1062 & $17:59:44.10$ & $-28:05:53.7$ &         $16.52749$ &              $5383.916$ &         $0.327$ &  $0.00445$ &    15.152 &    16.495 &  7.76 &              1.00 \\
OGLE-TR-1063 & $17:59:51.37$ & $-29:17:22.5$ &        $3.5835181$ &             $5261.3897$ &        $0.1596$ &  $0.01070$ &    14.703 &    15.560 & 51.49 &              1.00 \\
OGLE-TR-1064 & $18:00:00.26$ & $-29:15:02.6$ &        $0.6700153$ &             $5286.4298$ &        $0.0945$ &  $0.00320$ &    15.472 &    16.582 & 20.52 &              0.99 \\
OGLE-TR-1065 & $18:00:17.82$ & $-28:49:53.7$ &         $4.967119$ &             $5380.6973$ &         $0.194$ &  $0.00310$ &    15.463 &    17.413 & 11.33 &              1.00 \\
OGLE-TR-1066 & $18:00:24.34$ & $-28:20:56.8$ &         $10.03302$ &              $5386.199$ &         $0.588$ &  $0.00219$ &    15.257 &    16.412 & 11.24 &              0.84 \\
OGLE-TR-1067 & $18:01:02.46$ & $-29:09:54.7$ &         $2.573161$ &              $5376.738$ &         $0.325$ &  $0.00217$ &    14.744 &    16.651 & 23.79 &              0.94 \\
OGLE-TR-1068 & $18:01:11.74$ & $-28:53:43.1$ &       $0.68587341$ &             $5376.9436$ &        $0.0858$ &  $0.00619$ &    12.800 &    13.359 & 41.69 &              1.00 \\
OGLE-TR-1069 & $18:01:16.03$ & $-28:25:46.1$ &        $10.605366$ &             $5383.6787$ &        $0.5106$ &  $0.01568$ &    15.339 &    16.611 & 54.25 &              1.00 \\
OGLE-TR-1070 & $18:01:17.40$ & $-28:30:28.0$ &        $1.7293835$ &             $5380.2336$ &        $0.1135$ &  $0.01428$ &    15.400 &    16.324 & 34.25 &              0.81 \\
OGLE-TR-1071 & $18:01:17.69$ & $-28:48:23.4$ &          $4.26703$ &              $5383.137$ &         $0.458$ &   $0.0049$ &    15.424 &    17.366 & 15.87 &              1.00 \\
OGLE-TR-1072 & $18:01:20.90$ & $-28:26:24.4$ &        $1.6133903$ &             $5377.4958$ &        $0.0986$ &  $0.00808$ &    15.112 &    16.169 & 33.32 &              1.00 \\
OGLE-TR-1073 & $18:01:28.64$ & $-28:25:28.0$ &          $6.92344$ &              $5380.819$ &         $0.400$ &  $0.00130$ &    15.283 &    17.176 &  6.60 &              1.00 \\
OGLE-TR-1074 & $18:01:46.42$ & $-28:00:52.9$ &        $0.7182068$ &             $5376.8768$ &        $0.0807$ &  $0.00344$ &    15.167 &    16.364 & 24.77 &              0.92 \\
OGLE-TR-1075 & $18:01:47.87$ & $-28:01:19.6$ &         $30.55566$ &             $5390.5423$ &         $0.441$ &  $0.01590$ &    15.297 &    16.490 & 39.16 &              1.00 \\
OGLE-TR-1076 & $18:01:59.91$ & $-28:46:32.9$ &         $5.785527$ &             $5380.2351$ &        $0.2844$ &  $0.01468$ &    15.057 &    16.128 & 25.86 &              1.00 \\
OGLE-TR-1077 & $18:02:01.98$ & $-28:43:15.8$ &         $14.40827$ &              $5381.544$ &         $0.255$ &  $0.00397$ &    14.606 &    15.541 &  6.52 &              0.99 \\
OGLE-TR-1078 & $18:02:16.54$ & $-29:05:43.6$ &        $4.3297076$ &             $5377.6280$ &        $0.2245$ &  $0.00740$ &    14.700 &    15.665 & 45.05 &              1.00 \\
OGLE-TR-1079 & $18:02:44.77$ & $-28:48:25.6$ &         $13.45047$ &              $5381.447$ &         $0.554$ &  $0.00294$ &    15.351 &    17.201 &  7.58 &              1.00 \\
OGLE-TR-1080 & $18:02:50.21$ & $-28:21:44.1$ &  $4.7576944$ &             $5377.9397$ &        $0.2339$ &  $0.00446$ &    14.871 &    16.122 & 25.76 &  1.00 \\
OGLE-TR-1081 & $18:03:14.00$ & $-28:27:14.8$ &         $6.406942$ &             $5377.5162$ &        $0.1741$ &  $0.00608$ &    14.978 &    15.934 & 23.62 &              1.00 \\
OGLE-TR-1082 & $18:03:15.26$ & $-28:45:00.7$ &         $3.403645$ &             $5379.0038$ &         $0.379$ &  $0.00377$ &    15.398 &    17.377 & 18.94 &              1.00 \\
OGLE-TR-1083 & $18:03:15.36$ & $-28:24:25.6$ &         $5.410081$ &             $5380.6355$ &        $0.1406$ &  $0.01005$ &    15.057 &    16.084 & 20.41 &              1.00 \\
OGLE-TR-1084 & $18:03:18.58$ & $-28:24:45.1$ &         $39.79726$ &             $5402.3122$ &        $0.2763$ &  $0.01558$ &    14.141 &    14.969 & 39.18 &              0.94 \\
OGLE-TR-1085 & $18:03:21.39$ & $-28:12:40.2$ &         $3.636724$ &             $5377.8549$ &         $0.129$ &  $0.00316$ &    15.362 &    17.017 & 12.74 &              1.00 \\
OGLE-TR-1086 & $18:03:39.15$ & $-28:47:04.6$ &         $6.703196$ &              $5383.100$ &         $0.237$ &  $0.00510$ &    15.324 &    16.785 & 17.18 &              0.94 \\
OGLE-TR-1087 & $18:03:46.16$ & $-28:07:56.2$ &         $4.243165$ &             $5378.3633$ &         $0.206$ &  $0.00654$ &    15.148 &    16.438 & 30.74 &              1.00 \\
OGLE-TR-1088 & $18:03:47.49$ & $-28:05:03.0$ &        $5.0748842$ &            $5378.05549$ &        $0.1469$ &  $0.01584$ &    14.438 &    15.373 & 66.06 &              1.00 \\
OGLE-TR-1089 & $18:04:07.51$ & $-28:41:30.5$ &         $51.09021$ &              $5384.643$ &         $0.346$ &   $0.0186$ &    14.961 &    16.088 & 15.94 &              1.00 \\
OGLE-TR-1090 & $18:04:14.72$ & $-28:20:22.9$ &         $6.555184$ &             $5376.6412$ &        $0.2352$ &  $0.00525$ &    14.240 &    14.918 & 18.22 &              1.00 \\
OGLE-TR-1091 & $18:04:18.26$ & $-28:24:30.5$ &         $55.74533$ &              $5427.429$ &         $0.429$ &   $0.0129$ &    15.249 &    17.054 & 21.54 &              0.99 \\
OGLE-TR-1092 & $18:04:23.84$ & $-28:48:50.7$ &         $3.660176$ &              $5380.164$ &         $0.212$ &  $0.00188$ &    15.294 &    16.873 &  7.65 &              1.00 \\
OGLE-TR-1093 & $18:04:46.65$ & $-28:57:20.9$ &        $1.4751354$ &             $5377.1974$ &         $0.134$ &   $0.0113$ &    14.756 &    16.523 & 80.19 &              0.98 \\
OGLE-TR-1094 & $18:04:49.42$ & $-28:00:44.0$ &         $13.20038$ &              $5388.402$ &         $0.312$ &   $0.0102$ &    15.497 &    17.402 &  9.66 &              1.00 \\
OGLE-TR-1095 & $18:04:53.49$ & $-28:13:33.0$ &        $0.7848204$ &             $5377.2801$ &        $0.0864$ &  $0.00282$ &    15.239 &    17.210 & 15.83 &              0.90 \\
OGLE-TR-1096 & $18:05:07.41$ & $-28:15:56.6$ &         $3.437753$ &             $5377.2231$ &         $0.446$ &  $0.00514$ &    15.446 &    17.249 & 27.43 &              1.00 \\
OGLE-TR-1097 & $18:05:15.35$ & $-28:08:47.3$ &        $10.756626$ &             $5385.3247$ &        $0.1199$ &  $0.01424$ &    15.128 &    16.745 & 29.74 &              0.90 \\
OGLE-TR-1098 & $18:05:29.39$ & $-28:15:28.8$ &         $5.880841$ &             $5379.1189$ &        $0.1454$ &   $0.0203$ &    15.372 &    16.404 & 40.22 &              0.93 \\
OGLE-TR-1099 & $18:05:36.98$ & $-28:28:27.6$ &         $10.46173$ &              $5376.945$ &         $0.630$ &  $0.00279$ &    15.369 &    17.213 & 14.37 &              1.00 \\
\end{supertabular}}
\medskip
\end{landscape}

\newpage 

\begin{landscape}
{
\TabCapp{12pt}{ Selected physical parameters of systems with planetary candidates.}
\footnotesize
\tablefirsthead{
\hline\noalign{\smallskip}
 Name &  $M_{*}$ &  $R_{*} $ & $\rm{log}g$ &$T_{\rm{eff}}$   &   $\rm{[Fe/H]}$ &                $d$   &    $a$   &     $i$ &   $T_{\rm{eq}}$ &  $R_{\rm{P}}$  \\
 &    [$M_{\odot}$] &            [$R_{\odot}$] &  [cgs] &[K] &       $
[dex] $ &  [pc] & [AU] &   [deg] &    [K] &  [$R_{\rm{J}}$] \\\noalign{\smallskip}
\hline\noalign{\smallskip} }
\tablehead{\multicolumn{11}{c}{ \normalsize	 {\spaceskip 2pt plus 1pt minus 1pt T a b l e  4}
}\\\noalign{\smallskip} \multicolumn{11}{c}{ \normalsize	{ Continued }} \\\noalign{\bigskip}
\hline\noalign{\smallskip}
 Name &  $M_{*} $ &  $R_{*} $ & $\rm{log}g$ &$T_{\rm{eff}}$   &   $\rm{[Fe/H]}$ &                $d$   &    $a$   &     $i$ &   $T_{\rm{eq}}$ &  $R_{\rm{P}}$  \\\rule{0pt}{0pt}
 &    [$M_{\odot}$] &            [$R_{\odot}$] &  [cgs] &[K] &       $
[dex] $ &  [pc] & [AU] &   [deg] &    [K] &  [$R_{\rm{J}}$] \\\noalign{\smallskip}
\hline
\noalign{\smallskip}
}
\setlength\tabcolsep{4pt}
\tablelasttail{\noalign{\smallskip}\hline}
\renewcommand{\arraystretch}{1.5}
\begin{supertabular}{ccccccccccc}
OGLE-TR-1001 &        $2.44^{+0.42}_{-0.36}$ &     $1.57^{+0.23}_{-0.14}$ &     $4.42^{+0.13}_{-0.15}$ &     $14400^{+2500}_{-3800}$ &        $-1.5^{+1.5}_{-1.6}$ &      $1810^{+91}_{-79}$ &     $0.0437^{+0.0023}_{-0.0024}$ &        $84.2^{+2.1}_{-2.3}$ &       $4190^{+480}_{-860}$ &     $1.39^{+0.22}_{-0.14}$ \\
OGLE-TR-1002 &        $1.27^{+0.12}_{-0.14}$ &  $1.429^{+0.086}_{-0.099}$ &  $4.232^{+0.062}_{-0.059}$ &        $6030^{+310}_{-280}$ &      $0.38^{+0.11}_{-0.18}$ &    $1660^{+130}_{-120}$ &     $0.1184^{+0.0036}_{-0.0044}$ &     $87.28^{+0.26}_{-0.22}$ &         $1007^{+51}_{-46}$ &     $1.66^{+0.12}_{-0.15}$ \\
OGLE-TR-1003 &     $0.927^{+0.130}_{-0.090}$ &  $1.021^{+0.098}_{-0.075}$ &  $4.393^{+0.051}_{-0.067}$ &        $5770^{+310}_{-250}$ &     $-0.11^{+0.23}_{-0.40}$ &     $1088^{+120}_{-93}$ &     $0.0377^{+0.0017}_{-0.0013}$ &        $88.2^{+1.2}_{-1.7}$ &         $1452^{+96}_{-78}$ &  $0.529^{+0.057}_{-0.042}$ \\
OGLE-TR-1004 &        $1.93^{+0.14}_{-0.14}$ &     $1.96^{+0.19}_{-0.18}$ &  $4.142^{+0.077}_{-0.079}$ &        $7780^{+400}_{-400}$ &      $0.31^{+0.12}_{-0.19}$ &    $1690^{+140}_{-130}$ &     $0.1166^{+0.0029}_{-0.0033}$ &     $86.64^{+0.46}_{-0.51}$ &         $1536^{+76}_{-73}$ &     $1.52^{+0.16}_{-0.16}$ \\
OGLE-TR-1005 &     $0.965^{+0.330}_{-0.075}$ &  $1.763^{+0.210}_{-0.090}$ &  $3.949^{+0.052}_{-0.062}$ &        $7800^{+830}_{-470}$ &     $-3.99^{+1.20}_{-0.31}$ &    $2210^{+260}_{-130}$ &     $0.0533^{+0.0055}_{-0.0014}$ &        $87.9^{+1.5}_{-1.8}$ &       $2160^{+230}_{-120}$ &  $1.389^{+0.180}_{-0.078}$ \\
OGLE-TR-1006 &        $1.74^{+0.14}_{-0.17}$ &     $2.19^{+0.20}_{-0.15}$ &  $4.003^{+0.044}_{-0.086}$ &        $5310^{+300}_{-310}$ &      $0.30^{+0.21}_{-0.26}$ &     $1520^{+120}_{-97}$ &  $0.03521^{+0.00090}_{-0.00110}$ &        $75.1^{+1.1}_{-2.0}$ &        $2022^{+100}_{-84}$ &  $1.131^{+0.150}_{-0.088}$ \\
OGLE-TR-1007 &        $1.65^{+0.21}_{-0.12}$ &     $2.95^{+0.24}_{-0.23}$ &  $3.726^{+0.053}_{-0.053}$ &        $6800^{+580}_{-520}$ &     $-3.32^{+1.50}_{-0.49}$ &    $1690^{+170}_{-160}$ &     $0.0569^{+0.0020}_{-0.0015}$ &     $78.22^{+0.96}_{-1.00}$ &       $2350^{+210}_{-180}$ &     $2.98^{+0.29}_{-0.27}$ \\
OGLE-TR-1008 &        $0.98^{+0.29}_{-0.35}$ &     $2.10^{+0.19}_{-0.16}$ &     $3.78^{+0.15}_{-0.20}$ &        $5130^{+430}_{-390}$ &        $-2.3^{+2.0}_{-1.4}$ &    $1770^{+130}_{-120}$ &        $0.312^{+0.028}_{-0.041}$ &     $88.87^{+0.49}_{-0.44}$ &          $652^{+51}_{-56}$ &  $1.170^{+0.110}_{-0.087}$ \\
OGLE-TR-1009 &        $4.19^{+0.56}_{-0.52}$ &  $2.185^{+0.098}_{-0.094}$ &  $4.382^{+0.027}_{-0.030}$ &     $21600^{+1800}_{-2100}$ &     $-3.13^{+1.00}_{-0.57}$ &    $3160^{+300}_{-280}$ &        $0.302^{+0.013}_{-0.013}$ &     $89.80^{+0.14}_{-0.21}$ &       $2800^{+230}_{-260}$ &  $2.218^{+0.099}_{-0.096}$ \\
OGLE-TR-1010 &     $0.826^{+0.039}_{-0.039}$ &  $1.711^{+0.140}_{-0.095}$ &  $3.888^{+0.056}_{-0.072}$ &        $6900^{+260}_{-270}$ &  $-4.437^{+0.130}_{-0.076}$ &      $1695^{+96}_{-80}$ &  $0.03116^{+0.00048}_{-0.00050}$ &        $83.9^{+3.0}_{-2.8}$ &         $2470^{+88}_{-81}$ &  $1.390^{+0.140}_{-0.092}$ \\
OGLE-TR-1011 &        $1.56^{+0.22}_{-0.17}$ &     $1.94^{+0.42}_{-0.27}$ &  $4.070^{+0.086}_{-0.160}$ &        $5150^{+380}_{-300}$ &      $0.31^{+0.20}_{-0.23}$ &    $1870^{+310}_{-210}$ &     $0.1525^{+0.0080}_{-0.0075}$ &        $88.6^{+1.0}_{-1.3}$ &          $896^{+62}_{-51}$ &     $1.52^{+0.42}_{-0.25}$ \\
OGLE-TR-1012 &        $1.88^{+0.19}_{-0.25}$ &     $1.78^{+0.15}_{-0.12}$ &  $4.205^{+0.056}_{-0.088}$ &        $8170^{+450}_{-400}$ &      $0.10^{+0.25}_{-0.49}$ &    $2490^{+260}_{-220}$ &     $0.0569^{+0.0018}_{-0.0025}$ &        $86.4^{+1.8}_{-1.6}$ &       $2220^{+140}_{-110}$ &  $1.088^{+0.099}_{-0.075}$ \\
OGLE-TR-1013 &        $2.00^{+0.60}_{-0.55}$ &     $1.47^{+0.13}_{-0.12}$ &     $4.40^{+0.13}_{-0.15}$ &     $12100^{+3200}_{-2100}$ &     $-1.59^{+0.93}_{-1.50}$ &    $2670^{+350}_{-260}$ &        $0.206^{+0.019}_{-0.021}$ &     $89.05^{+0.56}_{-0.48}$ &       $1570^{+310}_{-220}$ &  $1.049^{+0.085}_{-0.075}$ \\
OGLE-TR-1014 &        $0.93^{+0.21}_{-0.11}$ &     $1.93^{+0.15}_{-0.14}$ &  $3.842^{+0.110}_{-0.086}$ &        $6020^{+640}_{-560}$ &     $-0.76^{+0.47}_{-1.20}$ &      $1100^{+67}_{-60}$ &     $0.1023^{+0.0068}_{-0.0040}$ &        $88.1^{+1.3}_{-1.2}$ &        $1252^{+110}_{-91}$ &  $0.944^{+0.070}_{-0.067}$ \\
OGLE-TR-1015 &     $0.916^{+0.160}_{-0.094}$ &     $2.26^{+0.28}_{-0.21}$ &  $3.696^{+0.094}_{-0.100}$ &        $5530^{+520}_{-370}$ &     $-0.69^{+0.38}_{-0.87}$ &    $2690^{+350}_{-270}$ &     $0.1250^{+0.0067}_{-0.0044}$ &        $88.5^{+1.0}_{-1.2}$ &        $1143^{+100}_{-81}$ &  $1.071^{+0.098}_{-0.076}$ \\
OGLE-TR-1016 &        $1.01^{+0.42}_{-0.18}$ &     $2.32^{+0.29}_{-0.31}$ &     $3.72^{+0.21}_{-0.14}$ &        $5620^{+650}_{-510}$ &     $-0.29^{+0.42}_{-1.20}$ &    $2680^{+320}_{-290}$ &     $0.0495^{+0.0058}_{-0.0029}$ &        $79.9^{+2.0}_{-1.9}$ &       $1830^{+190}_{-130}$ &  $0.983^{+0.088}_{-0.084}$ \\
OGLE-TR-1017 &        $1.90^{+0.91}_{-0.50}$ &     $1.28^{+0.44}_{-0.16}$ &     $4.49^{+0.23}_{-0.29}$ &     $12700^{+5200}_{-5100}$ &        $-2.6^{+2.7}_{-1.1}$ &    $2860^{+630}_{-570}$ &        $0.100^{+0.014}_{-0.010}$ &        $87.9^{+1.5}_{-1.9}$ &       $2200^{+650}_{-690}$ &     $1.34^{+0.62}_{-0.18}$ \\
OGLE-TR-1018 &        $1.24^{+0.25}_{-0.28}$ &     $1.57^{+0.14}_{-0.12}$ &  $4.134^{+0.087}_{-0.120}$ &        $6360^{+820}_{-630}$ &      $0.04^{+0.32}_{-1.20}$ &    $1890^{+240}_{-210}$ &     $0.0517^{+0.0031}_{-0.0041}$ &        $86.2^{+1.9}_{-1.7}$ &       $1690^{+240}_{-160}$ &  $1.021^{+0.093}_{-0.080}$ \\
OGLE-TR-1019 &     $0.820^{+0.078}_{-0.048}$ &  $1.385^{+0.079}_{-0.069}$ &  $4.076^{+0.049}_{-0.051}$ &        $6970^{+410}_{-330}$ &     $-4.28^{+1.20}_{-0.24}$ &      $1219^{+89}_{-80}$ &     $0.1485^{+0.0045}_{-0.0029}$ &     $89.65^{+0.24}_{-0.35}$ &         $1026^{+58}_{-51}$ &  $1.975^{+0.100}_{-0.092}$ \\
OGLE-TR-1020 &        $0.57^{+0.17}_{-0.28}$ &  $0.865^{+0.055}_{-0.041}$ &     $4.29^{+0.12}_{-0.25}$ &        $3710^{+190}_{-110}$ &      $0.16^{+0.24}_{-1.60}$ &       $411^{+29}_{-26}$ &        $0.182^{+0.017}_{-0.038}$ &  $89.858^{+0.098}_{-0.330}$ &          $403^{+34}_{-30}$ &  $1.206^{+0.140}_{-0.085}$ \\
OGLE-TR-1021 &     $1.504^{+0.100}_{-0.085}$ &  $1.459^{+0.062}_{-0.059}$ &  $4.288^{+0.029}_{-0.029}$ &        $6990^{+360}_{-250}$ &      $0.17^{+0.18}_{-0.14}$ &      $1586^{+85}_{-77}$ &     $0.3053^{+0.0067}_{-0.0058}$ &  $89.199^{+0.074}_{-0.072}$ &          $737^{+36}_{-28}$ &     $2.20^{+0.11}_{-0.10}$ \\
OGLE-TR-1022 &     $0.871^{+0.064}_{-0.063}$ &     $2.59^{+0.69}_{-0.32}$ &     $3.56^{+0.11}_{-0.23}$ &        $6490^{+450}_{-790}$ &     $-4.16^{+0.96}_{-0.17}$ &    $1910^{+210}_{-140}$ &     $0.0778^{+0.0025}_{-0.0024}$ &        $84.1^{+1.7}_{-2.9}$ &         $1801^{+69}_{-74}$ &     $1.45^{+0.47}_{-0.22}$ \\
OGLE-TR-1023 &     $1.331^{+0.091}_{-0.110}$ &  $1.386^{+0.071}_{-0.073}$ &  $4.279^{+0.040}_{-0.046}$ &        $6290^{+250}_{-240}$ &      $0.34^{+0.14}_{-0.21}$ &      $1481^{+93}_{-87}$ &     $0.0757^{+0.0017}_{-0.0021}$ &     $85.92^{+0.29}_{-0.28}$ &         $1296^{+56}_{-51}$ &     $1.71^{+0.11}_{-0.12}$ \\
OGLE-TR-1024 &        $1.44^{+0.27}_{-1.00}$ &     $2.03^{+0.17}_{-0.15}$ &     $3.96^{+0.10}_{-0.48}$ &        $5190^{+430}_{-430}$ &     $-0.09^{+0.42}_{-3.20}$ &    $1630^{+180}_{-160}$ &     $0.0283^{+0.0016}_{-0.0089}$ &       $77.0^{+4.6}_{-11.0}$ &       $2230^{+210}_{-160}$ &  $1.029^{+0.086}_{-0.081}$ \\
OGLE-TR-1025 &        $0.42^{+1.10}_{-0.14}$ &     $2.85^{+0.23}_{-0.21}$ &     $3.16^{+0.50}_{-0.20}$ &        $4880^{+260}_{-240}$ &     $-3.34^{+3.30}_{-0.48}$ &    $1980^{+190}_{-150}$ &     $0.0550^{+0.0280}_{-0.0064}$ &        $79.2^{+5.8}_{-2.8}$ &       $1670^{+110}_{-180}$ &  $1.030^{+0.080}_{-0.074}$ \\
OGLE-TR-1026 &     $0.857^{+0.140}_{-0.067}$ &     $2.73^{+0.32}_{-0.28}$ &  $3.503^{+0.100}_{-0.082}$ &        $5760^{+480}_{-430}$ &     $-1.76^{+0.94}_{-1.20}$ &    $1710^{+240}_{-180}$ &     $0.0484^{+0.0026}_{-0.0015}$ &        $87.4^{+1.8}_{-3.0}$ &       $2100^{+170}_{-170}$ &  $1.196^{+0.140}_{-0.090}$ \\
OGLE-TR-1027 &        $2.06^{+0.47}_{-0.51}$ &     $2.95^{+1.10}_{-0.39}$ &     $3.80^{+0.11}_{-0.28}$ &        $5320^{+370}_{-290}$ &      $0.26^{+0.23}_{-0.35}$ &    $2570^{+780}_{-340}$ &     $0.0477^{+0.0042}_{-0.0037}$ &        $75.2^{+2.5}_{-7.7}$ &       $2090^{+240}_{-170}$ &     $1.21^{+1.20}_{-0.15}$ \\
OGLE-TR-1028 &        $1.11^{+0.19}_{-0.14}$ &     $2.06^{+0.23}_{-0.16}$ &  $3.854^{+0.085}_{-0.095}$ &        $5290^{+510}_{-330}$ &      $0.22^{+0.23}_{-0.26}$ &     $1130^{+130}_{-99}$ &  $0.01676^{+0.00090}_{-0.00074}$ &        $60.3^{+4.1}_{-5.2}$ &       $2830^{+240}_{-160}$ &     $1.40^{+0.22}_{-0.14}$ \\
OGLE-TR-1029 &     $1.025^{+0.075}_{-0.085}$ &  $0.995^{+0.050}_{-0.039}$ &  $4.453^{+0.029}_{-0.043}$ &        $5650^{+210}_{-200}$ &      $0.23^{+0.19}_{-0.19}$ &      $1183^{+78}_{-70}$ &  $0.03711^{+0.00088}_{-0.00100}$ &        $87.9^{+1.2}_{-1.1}$ &         $1414^{+62}_{-55}$ &  $1.067^{+0.065}_{-0.047}$ \\
OGLE-TR-1030 &        $1.40^{+0.11}_{-0.12}$ &  $1.556^{+0.160}_{-0.097}$ &  $4.204^{+0.058}_{-0.098}$ &        $6280^{+330}_{-330}$ &      $0.37^{+0.11}_{-0.17}$ &     $1536^{+120}_{-99}$ &     $0.1517^{+0.0039}_{-0.0045}$ &     $88.88^{+0.62}_{-0.65}$ &          $974^{+49}_{-41}$ &  $1.381^{+0.190}_{-0.099}$ \\
OGLE-TR-1031 &        $1.15^{+0.22}_{-0.17}$ &     $1.96^{+0.17}_{-0.15}$ &  $3.919^{+0.095}_{-0.110}$ &        $5490^{+330}_{-420}$ &      $0.18^{+0.23}_{-0.24}$ &    $1410^{+130}_{-110}$ &     $0.0374^{+0.0021}_{-0.0018}$ &        $85.5^{+3.0}_{-3.4}$ &         $1901^{+92}_{-84}$ &  $1.009^{+0.073}_{-0.068}$ \\
OGLE-TR-1032 &     $1.417^{+0.084}_{-0.051}$ &  $1.699^{+0.050}_{-0.040}$ &  $4.131^{+0.014}_{-0.015}$ &        $5860^{+290}_{-280}$ &      $0.45^{+0.11}_{-0.21}$ &      $1598^{+83}_{-74}$ &     $0.1287^{+0.0025}_{-0.0016}$ &     $89.65^{+0.25}_{-0.36}$ &         $1026^{+50}_{-49}$ &  $2.232^{+0.067}_{-0.050}$ \\
OGLE-TR-1033 &     $0.883^{+0.120}_{-0.083}$ &  $1.725^{+0.150}_{-0.082}$ &  $3.916^{+0.041}_{-0.062}$ &        $7090^{+550}_{-420}$ &     $-3.57^{+1.10}_{-0.42}$ &    $1880^{+160}_{-130}$ &     $0.1286^{+0.0055}_{-0.0041}$ &     $89.18^{+0.58}_{-0.82}$ &         $1258^{+92}_{-73}$ &  $1.381^{+0.130}_{-0.072}$ \\
OGLE-TR-1034 &     $0.782^{+0.032}_{-0.022}$ &  $1.434^{+0.061}_{-0.061}$ &  $4.022^{+0.041}_{-0.037}$ &        $6530^{+210}_{-200}$ &  $-4.480^{+0.080}_{-0.059}$ &      $1409^{+75}_{-72}$ &  $0.05554^{+0.00075}_{-0.00052}$ &     $85.05^{+0.44}_{-0.41}$ &         $1598^{+54}_{-53}$ &     $2.02^{+0.11}_{-0.10}$ \\
OGLE-TR-1035 &     $1.633^{+0.096}_{-0.092}$ &  $1.601^{+0.070}_{-0.067}$ &  $4.244^{+0.037}_{-0.042}$ &        $7230^{+330}_{-290}$ &      $0.15^{+0.16}_{-0.13}$ &      $1262^{+42}_{-39}$ &     $0.3344^{+0.0064}_{-0.0064}$ &  $88.950^{+0.067}_{-0.068}$ &          $763^{+29}_{-25}$ &     $1.73^{+0.11}_{-0.11}$ \\
OGLE-TR-1036 &     $0.794^{+0.083}_{-0.034}$ &  $1.329^{+0.050}_{-0.043}$ &  $4.099^{+0.028}_{-0.025}$ &        $6690^{+320}_{-380}$ &     $-3.51^{+2.10}_{-0.86}$ &      $1096^{+87}_{-78}$ &  $0.01739^{+0.00059}_{-0.00025}$ &     $71.70^{+0.71}_{-0.75}$ &       $2810^{+140}_{-170}$ &  $1.668^{+0.110}_{-0.091}$ \\
OGLE-TR-1037 &        $1.42^{+0.32}_{-0.59}$ &     $2.56^{+0.26}_{-0.22}$ &     $3.78^{+0.13}_{-0.27}$ &        $6090^{+440}_{-750}$ &      $0.00^{+0.31}_{-2.30}$ &    $2280^{+200}_{-170}$ &     $0.0778^{+0.0053}_{-0.0120}$ &        $83.3^{+1.6}_{-2.3}$ &       $1690^{+120}_{-110}$ &  $1.001^{+0.085}_{-0.080}$ \\
OGLE-TR-1038 &     $0.271^{+0.110}_{-0.073}$ &     $2.66^{+0.19}_{-0.18}$ &     $3.02^{+0.17}_{-0.15}$ &        $4630^{+230}_{-220}$ &     $-3.51^{+0.68}_{-0.35}$ &      $1187^{+41}_{-40}$ &     $0.0352^{+0.0031}_{-0.0028}$ &        $84.5^{+3.9}_{-5.0}$ &        $1936^{+100}_{-80}$ &     $1.79^{+0.22}_{-0.39}$ \\
OGLE-TR-1039 &        $1.63^{+0.18}_{-0.24}$ &     $2.12^{+0.21}_{-0.15}$ &  $3.992^{+0.084}_{-0.130}$ &        $6700^{+480}_{-430}$ &      $0.08^{+0.22}_{-0.21}$ &    $2320^{+220}_{-170}$ &     $0.1017^{+0.0037}_{-0.0052}$ &     $85.53^{+0.62}_{-0.91}$ &         $1485^{+91}_{-78}$ &  $1.167^{+0.150}_{-0.096}$ \\
OGLE-TR-1040 &        $1.00^{+0.25}_{-0.16}$ &  $0.993^{+0.079}_{-0.068}$ &  $4.450^{+0.100}_{-0.098}$ &       $7370^{+1700}_{-810}$ &        $-2.8^{+1.6}_{-1.2}$ &    $2010^{+350}_{-270}$ &     $0.1438^{+0.0110}_{-0.0081}$ &     $89.08^{+0.40}_{-0.33}$ &        $940^{+180}_{-110}$ &  $1.058^{+0.093}_{-0.078}$ \\
OGLE-TR-1041 &        $1.06^{+0.21}_{-0.14}$ &  $1.248^{+0.069}_{-0.061}$ &  $4.279^{+0.035}_{-0.043}$ &       $7190^{+790}_{-1100}$ &        $-1.4^{+1.6}_{-2.1}$ &      $1370^{+82}_{-76}$ &     $0.0423^{+0.0025}_{-0.0018}$ &     $88.73^{+0.88}_{-1.30}$ &       $1880^{+200}_{-290}$ &  $1.140^{+0.060}_{-0.054}$ \\
OGLE-TR-1042 &     $0.871^{+0.150}_{-0.092}$ &  $1.546^{+0.086}_{-0.071}$ &  $4.006^{+0.025}_{-0.026}$ &        $5980^{+420}_{-460}$ &     $-0.80^{+0.80}_{-1.90}$ &    $1390^{+120}_{-100}$ &     $0.0727^{+0.0039}_{-0.0027}$ &     $89.33^{+0.47}_{-0.73}$ &        $1327^{+94}_{-100}$ &  $1.662^{+0.087}_{-0.073}$ \\
OGLE-TR-1043 &        $1.26^{+0.18}_{-0.17}$ &     $1.86^{+0.43}_{-0.26}$ &     $3.99^{+0.14}_{-0.16}$ &        $5750^{+350}_{-380}$ &      $0.24^{+0.17}_{-0.22}$ &    $1670^{+320}_{-230}$ &     $0.0361^{+0.0017}_{-0.0018}$ &        $78.2^{+2.1}_{-1.7}$ &       $1990^{+160}_{-130}$ &  $1.003^{+0.093}_{-0.084}$ \\
OGLE-TR-1044 &     $1.010^{+0.070}_{-0.069}$ &  $1.014^{+0.030}_{-0.029}$ &  $4.431^{+0.030}_{-0.033}$ &        $5580^{+130}_{-120}$ &      $0.34^{+0.14}_{-0.16}$ &       $804^{+22}_{-21}$ &     $0.0589^{+0.0013}_{-0.0014}$ &     $89.47^{+0.37}_{-0.53}$ &         $1117^{+23}_{-23}$ &  $2.015^{+0.066}_{-0.064}$ \\
OGLE-TR-1045 &     $0.797^{+0.074}_{-0.038}$ &  $1.317^{+0.052}_{-0.041}$ &  $4.106^{+0.025}_{-0.026}$ &        $7000^{+370}_{-300}$ &     $-4.18^{+0.99}_{-0.27}$ &    $1810^{+190}_{-190}$ &  $0.03010^{+0.00090}_{-0.00048}$ &     $88.97^{+0.75}_{-1.10}$ &        $2227^{+120}_{-91}$ &  $2.228^{+0.082}_{-0.070}$ \\
OGLE-TR-1046 &     $0.993^{+0.097}_{-0.079}$ &  $1.570^{+0.065}_{-0.064}$ &  $4.044^{+0.026}_{-0.029}$ &        $5560^{+380}_{-220}$ &      $0.05^{+0.23}_{-0.25}$ &      $1391^{+99}_{-84}$ &     $0.0943^{+0.0030}_{-0.0026}$ &     $89.47^{+0.37}_{-0.56}$ &         $1094^{+74}_{-43}$ &  $1.813^{+0.078}_{-0.071}$ \\
OGLE-TR-1047 &        $0.90^{+0.12}_{-0.18}$ &  $1.347^{+0.081}_{-0.085}$ &  $4.134^{+0.036}_{-0.064}$ &        $5530^{+270}_{-250}$ &     $-3.45^{+0.84}_{-0.38}$ &    $1720^{+140}_{-140}$ &     $0.0896^{+0.0038}_{-0.0064}$ &     $89.17^{+0.58}_{-0.82}$ &         $1039^{+52}_{-48}$ &  $1.519^{+0.100}_{-0.099}$ \\
OGLE-TR-1048 &        $0.85^{+0.21}_{-0.31}$ &     $1.15^{+0.16}_{-0.15}$ &  $4.230^{+0.062}_{-0.100}$ &       $5400^{+1100}_{-430}$ &        $-2.5^{+1.8}_{-1.1}$ &    $1530^{+240}_{-190}$ &     $0.0573^{+0.0054}_{-0.0084}$ &     $88.78^{+0.87}_{-1.40}$ &        $1186^{+220}_{-97}$ &     $1.47^{+0.22}_{-0.20}$ \\
OGLE-TR-1049 &        $0.71^{+0.47}_{-0.19}$ &     $3.40^{+0.33}_{-0.25}$ &     $3.22^{+0.22}_{-0.13}$ &        $5290^{+280}_{-330}$ &     $-2.63^{+1.20}_{-0.85}$ &    $2440^{+230}_{-180}$ &        $0.128^{+0.022}_{-0.012}$ &        $88.1^{+1.3}_{-1.9}$ &         $1306^{+77}_{-75}$ &  $1.144^{+0.120}_{-0.083}$ \\
OGLE-TR-1050 &     $0.122^{+0.061}_{-0.017}$ &  $0.953^{+0.069}_{-0.052}$ &  $3.579^{+0.130}_{-0.059}$ &        $4740^{+240}_{-200}$ &     $-3.01^{+0.81}_{-0.65}$ &      $962^{+100}_{-89}$ &  $0.01415^{+0.00170}_{-0.00089}$ &        $74.5^{+2.0}_{-1.5}$ &       $1860^{+110}_{-110}$ &  $1.152^{+0.093}_{-0.080}$ \\
OGLE-TR-1051 &     $0.946^{+0.120}_{-0.073}$ &     $4.00^{+0.21}_{-0.15}$ &  $3.210^{+0.031}_{-0.018}$ &        $4770^{+140}_{-150}$ &     $-0.20^{+0.23}_{-0.38}$ &    $2880^{+220}_{-180}$ &  $0.01956^{+0.00098}_{-0.00055}$ &       $78.8^{+7.8}_{-12.0}$ &       $3280^{+100}_{-110}$ &  $1.407^{+0.100}_{-0.069}$ \\
OGLE-TR-1052 &        $1.01^{+0.57}_{-0.21}$ &        $6.0^{+1.6}_{-1.9}$ &     $2.88^{+0.47}_{-0.20}$ &        $5360^{+320}_{-280}$ &        $-2.5^{+2.3}_{-1.1}$ &  $4900^{+1600}_{-1300}$ &     $0.0765^{+0.0084}_{-0.0051}$ &       $72.0^{+13.0}_{-6.2}$ &       $2260^{+290}_{-400}$ &     $2.44^{+0.76}_{-0.99}$ \\
OGLE-TR-1053 &        $1.11^{+0.41}_{-0.23}$ &     $1.53^{+0.27}_{-0.24}$ &  $4.128^{+0.110}_{-0.098}$ &        $7250^{+650}_{-530}$ &     $-1.10^{+0.86}_{-2.20}$ &    $2030^{+360}_{-260}$ &     $0.0493^{+0.0057}_{-0.0040}$ &        $82.8^{+1.2}_{-1.1}$ &       $1930^{+170}_{-130}$ &     $1.57^{+0.47}_{-0.31}$ \\
OGLE-TR-1054 &        $1.55^{+0.34}_{-0.27}$ &     $2.51^{+0.49}_{-0.30}$ &     $3.80^{+0.16}_{-0.15}$ &        $6060^{+940}_{-570}$ &        $-1.5^{+1.7}_{-1.7}$ &    $2670^{+360}_{-270}$ &     $0.0961^{+0.0065}_{-0.0061}$ &        $88.2^{+1.3}_{-2.4}$ &       $1540^{+140}_{-110}$ &     $1.22^{+0.41}_{-0.12}$ \\
OGLE-TR-1055 &        $0.97^{+0.21}_{-0.11}$ &     $3.00^{+0.33}_{-0.25}$ &  $3.477^{+0.093}_{-0.100}$ &        $5090^{+230}_{-190}$ &     $-0.46^{+0.23}_{-0.48}$ &    $1830^{+260}_{-200}$ &     $0.0323^{+0.0021}_{-0.0013}$ &        $72.5^{+4.0}_{-4.3}$ &       $2370^{+180}_{-130}$ &  $1.067^{+0.110}_{-0.089}$ \\
OGLE-TR-1056 &     $1.071^{+0.110}_{-0.066}$ &     $2.55^{+0.48}_{-0.38}$ &     $3.66^{+0.15}_{-0.13}$ &        $4690^{+160}_{-150}$ &      $0.41^{+0.13}_{-0.20}$ &    $1690^{+310}_{-220}$ &     $0.0791^{+0.0024}_{-0.0018}$ &        $82.3^{+1.2}_{-1.3}$ &        $1281^{+110}_{-99}$ &  $1.048^{+0.085}_{-0.078}$ \\
OGLE-TR-1057 &     $0.782^{+0.029}_{-0.014}$ &     $3.67^{+0.10}_{-0.10}$ &  $3.206^{+0.024}_{-0.023}$ &        $5590^{+220}_{-240}$ &     $-2.71^{+1.00}_{-0.91}$ &  $5400^{+1400}_{-1300}$ &  $0.05857^{+0.00071}_{-0.00034}$ &     $88.86^{+0.80}_{-1.20}$ &         $2130^{+89}_{-95}$ &  $2.498^{+0.084}_{-0.084}$ \\
OGLE-TR-1058 &        $1.82^{+0.16}_{-0.17}$ &     $2.62^{+0.36}_{-0.23}$ &  $3.854^{+0.082}_{-0.110}$ &        $6640^{+450}_{-360}$ &      $0.26^{+0.14}_{-0.20}$ &    $2460^{+280}_{-200}$ &     $0.1615^{+0.0052}_{-0.0051}$ &        $88.5^{+1.0}_{-1.1}$ &         $1299^{+81}_{-69}$ &  $1.171^{+0.170}_{-0.089}$ \\
OGLE-TR-1059 &        $1.02^{+0.31}_{-0.19}$ &     $1.48^{+0.13}_{-0.11}$ &     $4.11^{+0.12}_{-0.10}$ &        $6820^{+850}_{-610}$ &        $-1.4^{+1.4}_{-2.0}$ &    $1590^{+130}_{-110}$ &     $0.0741^{+0.0068}_{-0.0049}$ &        $88.3^{+1.1}_{-1.1}$ &  $1490.00^{+0.37}_{-0.19}$ &  $1.152^{+0.096}_{-0.077}$ \\
OGLE-TR-1060 &     $1.116^{+0.200}_{-0.089}$ &  $1.532^{+0.096}_{-0.093}$ &  $4.127^{+0.070}_{-0.073}$ &        $6190^{+560}_{-490}$ &     $-3.29^{+3.50}_{-0.53}$ &    $1810^{+140}_{-130}$ &     $0.1113^{+0.0064}_{-0.0030}$ &     $88.52^{+0.85}_{-0.64}$ &         $1100^{+91}_{-64}$ &     $1.83^{+0.14}_{-0.12}$ \\
OGLE-TR-1061 &        $1.79^{+0.24}_{-0.58}$ &  $1.805^{+0.086}_{-0.170}$ &  $4.173^{+0.031}_{-0.069}$ &       $8510^{+1100}_{-520}$ &     $-0.15^{+0.44}_{-2.90}$ &    $2090^{+140}_{-150}$ &     $0.0455^{+0.0020}_{-0.0056}$ &        $88.3^{+1.2}_{-1.8}$ &       $2600^{+340}_{-160}$ &  $1.678^{+0.081}_{-0.160}$ \\
OGLE-TR-1062 &        $1.01^{+0.14}_{-0.16}$ &     $1.70^{+0.16}_{-0.16}$ &  $3.981^{+0.110}_{-0.096}$ &        $6870^{+650}_{-900}$ &        $-1.6^{+1.5}_{-2.2}$ &    $1960^{+550}_{-350}$ &     $0.1289^{+0.0052}_{-0.0078}$ &     $89.07^{+0.64}_{-0.59}$ &       $1220^{+140}_{-180}$ &  $1.102^{+0.061}_{-0.068}$ \\
OGLE-TR-1063 &        $1.12^{+0.40}_{-0.13}$ &  $1.382^{+0.120}_{-0.082}$ &  $4.233^{+0.059}_{-0.076}$ &        $8290^{+440}_{-490}$ &        $-1.9^{+1.4}_{-1.6}$ &      $1527^{+89}_{-77}$ &     $0.0476^{+0.0051}_{-0.0019}$ &        $87.6^{+1.5}_{-1.5}$ &        $2137^{+99}_{-110}$ &  $1.392^{+0.130}_{-0.090}$ \\
OGLE-TR-1064 &        $1.09^{+0.27}_{-0.50}$ &     $1.87^{+0.14}_{-0.13}$ &     $3.92^{+0.11}_{-0.24}$ &        $5240^{+430}_{-310}$ &     $-1.22^{+0.76}_{-0.96}$ &    $2300^{+220}_{-200}$ &     $0.0156^{+0.0012}_{-0.0028}$ &       $61.9^{+5.7}_{-11.0}$ &       $2850^{+200}_{-200}$ &  $1.045^{+0.130}_{-0.097}$ \\
OGLE-TR-1065 &        $0.93^{+0.36}_{-0.26}$ &     $1.87^{+0.12}_{-0.14}$ &     $3.83^{+0.19}_{-0.12}$ &        $4270^{+240}_{-220}$ &      $0.25^{+0.24}_{-0.38}$ &    $1270^{+310}_{-140}$ &     $0.0564^{+0.0064}_{-0.0056}$ &        $83.8^{+1.4}_{-1.3}$ &         $1184^{+62}_{-52}$ &  $1.024^{+0.058}_{-0.092}$ \\
OGLE-TR-1066 &     $0.881^{+0.220}_{-0.088}$ &     $3.70^{+1.40}_{-0.77}$ &     $3.28^{+0.15}_{-0.25}$ &        $5430^{+370}_{-400}$ &        $-1.7^{+1.2}_{-1.6}$ &   $4540^{+1600}_{-880}$ &     $0.0917^{+0.0075}_{-0.0062}$ &        $86.8^{+2.4}_{-6.6}$ &       $1670^{+210}_{-150}$ &     $1.68^{+0.75}_{-0.39}$ \\
OGLE-TR-1067 &        $1.25^{+0.24}_{-0.19}$ &     $5.52^{+0.61}_{-0.75}$ &  $3.051^{+0.140}_{-0.099}$ &        $4620^{+140}_{-140}$ &      $0.36^{+0.16}_{-0.20}$ &    $2800^{+320}_{-360}$ &     $0.0417^{+0.0023}_{-0.0018}$ &        $56.0^{+7.1}_{-5.1}$ &       $2550^{+130}_{-160}$ &     $2.51^{+0.44}_{-0.49}$ \\
OGLE-TR-1068 &        $1.67^{+0.14}_{-0.15}$ &     $2.27^{+0.27}_{-0.43}$ &  $3.944^{+0.160}_{-0.089}$ &        $6820^{+360}_{-310}$ &      $0.09^{+0.19}_{-0.14}$ &    $1430^{+180}_{-200}$ &  $0.01856^{+0.00069}_{-0.00074}$ &        $58.6^{+7.2}_{-6.1}$ &       $3600^{+180}_{-190}$ &     $1.74^{+0.86}_{-0.44}$ \\
OGLE-TR-1069 &  $0.7726^{+0.0190}_{-0.0089}$ &  $2.506^{+0.037}_{-0.034}$ &  $3.530^{+0.011}_{-0.011}$ &        $5950^{+230}_{-260}$ &     $-2.96^{+0.97}_{-0.83}$ &    $3010^{+170}_{-160}$ &  $0.08670^{+0.00071}_{-0.00033}$ &     $89.62^{+0.27}_{-0.42}$ &         $1541^{+58}_{-67}$ &  $3.053^{+0.044}_{-0.041}$ \\
OGLE-TR-1070 &        $0.91^{+0.12}_{-0.15}$ &  $1.086^{+0.055}_{-0.060}$ &  $4.321^{+0.027}_{-0.027}$ &        $5970^{+790}_{-430}$ &     $-0.13^{+0.36}_{-3.60}$ &    $1680^{+120}_{-100}$ &     $0.0274^{+0.0011}_{-0.0015}$ &     $88.83^{+0.83}_{-1.30}$ &       $1820^{+240}_{-130}$ &  $1.259^{+0.063}_{-0.061}$ \\
OGLE-TR-1071 &        $1.36^{+0.74}_{-0.49}$ &        $7.7^{+1.9}_{-1.8}$ &     $2.81^{+0.17}_{-0.17}$ &        $4330^{+170}_{-170}$ &      $0.42^{+0.13}_{-0.24}$ &  $5500^{+1500}_{-1300}$ &     $0.0644^{+0.0097}_{-0.0079}$ &        $57.8^{+7.6}_{-6.1}$ &       $2270^{+180}_{-200}$ &        $5.5^{+3.3}_{-2.0}$ \\
OGLE-TR-1072 &        $0.97^{+0.16}_{-0.16}$ &  $1.140^{+0.068}_{-0.066}$ &  $4.311^{+0.075}_{-0.073}$ &        $6180^{+530}_{-370}$ &     $-0.28^{+0.37}_{-2.00}$ &      $1491^{+96}_{-85}$ &     $0.0271^{+0.0014}_{-0.0015}$ &        $84.4^{+1.8}_{-1.5}$ &       $1930^{+180}_{-130}$ &  $0.996^{+0.068}_{-0.065}$ \\
OGLE-TR-1073 &        $1.34^{+0.54}_{-0.49}$ &     $3.26^{+0.82}_{-0.36}$ &     $3.51^{+0.16}_{-0.30}$ &        $4550^{+170}_{-230}$ &      $0.34^{+0.18}_{-0.29}$ &    $2240^{+570}_{-300}$ &     $0.0795^{+0.0093}_{-0.0110}$ &        $84.9^{+3.4}_{-6.3}$ &        $1422^{+170}_{-85}$ &  $1.110^{+0.310}_{-0.092}$ \\
OGLE-TR-1074 &        $1.56^{+0.28}_{-0.25}$ &     $2.31^{+0.38}_{-0.24}$ &     $3.88^{+0.13}_{-0.12}$ &        $6180^{+690}_{-540}$ &        $-0.7^{+1.1}_{-2.6}$ &    $2860^{+320}_{-290}$ &  $0.01868^{+0.00097}_{-0.00130}$ &        $56.2^{+5.8}_{-7.8}$ &       $3350^{+270}_{-220}$ &     $1.29^{+1.10}_{-0.19}$ \\
OGLE-TR-1075 &     $0.790^{+0.050}_{-0.026}$ &  $1.578^{+0.079}_{-0.058}$ &  $3.946^{+0.033}_{-0.043}$ &        $6680^{+290}_{-250}$ &     $-4.14^{+0.54}_{-0.12}$ &    $2250^{+130}_{-120}$ &     $0.1768^{+0.0036}_{-0.0020}$ &     $89.47^{+0.36}_{-0.42}$ &          $962^{+41}_{-37}$ &  $1.931^{+0.110}_{-0.073}$ \\
OGLE-TR-1076 &        $0.63^{+0.13}_{-0.10}$ &  $1.625^{+0.093}_{-0.084}$ &  $3.819^{+0.044}_{-0.054}$ &        $5380^{+180}_{-230}$ &     $-3.45^{+0.80}_{-0.38}$ &    $1730^{+120}_{-110}$ &     $0.0540^{+0.0035}_{-0.0031}$ &     $88.73^{+0.90}_{-1.30}$ &         $1420^{+51}_{-60}$ &     $1.92^{+0.11}_{-0.10}$ \\
OGLE-TR-1077 &        $1.36^{+0.18}_{-0.32}$ &     $1.67^{+0.15}_{-0.13}$ &  $4.116^{+0.092}_{-0.140}$ &        $6370^{+430}_{-420}$ &      $0.04^{+0.24}_{-0.21}$ &    $1860^{+160}_{-140}$ &     $0.1296^{+0.0054}_{-0.0110}$ &     $88.13^{+0.90}_{-0.72}$ &         $1105^{+74}_{-57}$ &  $1.027^{+0.078}_{-0.072}$ \\
OGLE-TR-1078 &        $0.94^{+0.48}_{-0.19}$ &     $1.67^{+0.20}_{-0.11}$ &  $3.973^{+0.060}_{-0.056}$ &        $5510^{+250}_{-220}$ &     $-3.12^{+3.00}_{-0.64}$ &    $1850^{+240}_{-140}$ &     $0.0510^{+0.0075}_{-0.0037}$ &     $88.73^{+0.88}_{-1.50}$ &         $1521^{+67}_{-60}$ &  $1.397^{+0.160}_{-0.090}$ \\
OGLE-TR-1079 &        $0.46^{+0.36}_{-0.15}$ &     $2.95^{+1.10}_{-0.60}$ &     $3.17^{+0.25}_{-0.30}$ &        $4280^{+250}_{-270}$ &     $-0.56^{+0.46}_{-0.90}$ &    $2050^{+780}_{-400}$ &        $0.093^{+0.017}_{-0.012}$ &        $85.1^{+3.4}_{-4.1}$ &       $1170^{+150}_{-110}$ &     $1.56^{+0.86}_{-0.43}$ \\
OGLE-TR-1080 &     $1.599^{+0.130}_{-0.073}$ &  $2.057^{+0.120}_{-0.077}$ &  $4.018^{+0.025}_{-0.036}$ &        $6230^{+320}_{-300}$ &      $0.40^{+0.14}_{-0.22}$ &    $2340^{+180}_{-140}$ &  $0.06480^{+0.00170}_{-0.00098}$ &        $88.4^{+1.1}_{-1.5}$ &         $1697^{+85}_{-81}$ &  $1.334^{+0.083}_{-0.054}$ \\
OGLE-TR-1081 &        $1.21^{+0.12}_{-0.15}$ &  $1.397^{+0.087}_{-0.081}$ &  $4.227^{+0.068}_{-0.075}$ &        $5960^{+270}_{-250}$ &      $0.25^{+0.18}_{-0.20}$ &    $1880^{+130}_{-110}$ &     $0.0727^{+0.0024}_{-0.0031}$ &     $87.38^{+0.78}_{-0.72}$ &         $1261^{+51}_{-46}$ &  $1.059^{+0.078}_{-0.071}$ \\
OGLE-TR-1082 &     $1.034^{+0.074}_{-0.051}$ &     $3.15^{+0.13}_{-0.13}$ &  $3.459^{+0.032}_{-0.031}$ &        $4550^{+140}_{-110}$ &      $0.38^{+0.16}_{-0.30}$ &    $2560^{+210}_{-160}$ &  $0.04479^{+0.00100}_{-0.00076}$ &        $88.0^{+1.4}_{-2.1}$ &         $1838^{+63}_{-53}$ &  $1.877^{+0.081}_{-0.077}$ \\
OGLE-TR-1083 &     $1.067^{+0.090}_{-0.093}$ &  $1.112^{+0.062}_{-0.055}$ &  $4.375^{+0.051}_{-0.058}$ &        $5670^{+190}_{-180}$ &      $0.33^{+0.14}_{-0.18}$ &      $1468^{+87}_{-78}$ &     $0.0624^{+0.0017}_{-0.0019}$ &     $87.62^{+0.70}_{-0.60}$ &         $1155^{+44}_{-40}$ &  $1.084^{+0.077}_{-0.068}$ \\
OGLE-TR-1084 &     $1.231^{+0.076}_{-0.097}$ &  $1.257^{+0.066}_{-0.063}$ &  $4.330^{+0.041}_{-0.053}$ &        $6200^{+180}_{-180}$ &      $0.23^{+0.18}_{-0.17}$ &      $1320^{+57}_{-52}$ &     $0.2445^{+0.0049}_{-0.0066}$ &     $89.10^{+0.11}_{-0.12}$ &          $677^{+20}_{-19}$ &  $1.526^{+0.100}_{-0.095}$ \\
OGLE-TR-1085 &     $0.924^{+0.042}_{-0.400}$ &     $1.71^{+0.23}_{-0.15}$ &  $3.938^{+0.079}_{-0.370}$ &        $5220^{+150}_{-350}$ &     $-0.19^{+0.15}_{-2.50}$ &    $1610^{+130}_{-130}$ &  $0.04551^{+0.00091}_{-0.00730}$ &     $81.49^{+0.56}_{-3.70}$ &        $1573^{+96}_{-130}$ &  $0.942^{+0.076}_{-0.044}$ \\
OGLE-TR-1086 &     $1.004^{+0.120}_{-0.090}$ &     $2.07^{+1.70}_{-0.47}$ &     $3.81^{+0.22}_{-0.51}$ &        $5010^{+210}_{-200}$ &      $0.00^{+0.38}_{-0.45}$ &   $1710^{+1200}_{-350}$ &     $0.0720^{+0.0058}_{-0.0023}$ &        $84.5^{+3.0}_{-6.8}$ &       $1310^{+340}_{-140}$ &     $1.42^{+1.90}_{-0.38}$ \\
OGLE-TR-1087 &        $1.21^{+0.24}_{-0.25}$ &     $2.27^{+0.43}_{-0.66}$ &     $3.83^{+0.26}_{-0.19}$ &        $5640^{+860}_{-730}$ &      $0.17^{+0.26}_{-1.00}$ &    $2400^{+410}_{-430}$ &     $0.0565^{+0.0033}_{-0.0041}$ &        $82.6^{+4.8}_{-2.8}$ &       $1660^{+190}_{-120}$ &     $1.79^{+0.44}_{-0.59}$ \\
OGLE-TR-1088 &     $0.860^{+0.064}_{-0.084}$ &  $0.972^{+0.030}_{-0.035}$ &  $4.398^{+0.019}_{-0.022}$ &        $5670^{+620}_{-230}$ &     $-0.17^{+0.17}_{-1.60}$ &       $954^{+33}_{-31}$ &     $0.0553^{+0.0014}_{-0.0017}$ &     $89.40^{+0.42}_{-0.58}$ &        $1145^{+120}_{-46}$ &  $1.187^{+0.040}_{-0.037}$ \\
OGLE-TR-1089 &     $1.259^{+0.089}_{-0.100}$ &  $1.294^{+0.087}_{-0.086}$ &  $4.317^{+0.049}_{-0.059}$ &        $6050^{+240}_{-240}$ &   $0.434^{+0.092}_{-0.160}$ &    $1480^{+110}_{-110}$ &     $0.2910^{+0.0067}_{-0.0080}$ &     $89.42^{+0.21}_{-0.16}$ &          $614^{+29}_{-27}$ &     $1.72^{+0.13}_{-0.14}$ \\
OGLE-TR-1090 &        $1.38^{+0.16}_{-0.26}$ &     $1.83^{+0.25}_{-0.11}$ &  $4.044^{+0.072}_{-0.150}$ &        $6490^{+340}_{-320}$ &     $-0.12^{+0.16}_{-0.47}$ &    $2130^{+180}_{-140}$ &     $0.0771^{+0.0031}_{-0.0048}$ &        $87.7^{+1.5}_{-2.0}$ &         $1548^{+85}_{-66}$ &  $1.283^{+0.230}_{-0.091}$ \\
OGLE-TR-1091 &     $0.146^{+0.100}_{-0.036}$ &  $0.971^{+0.082}_{-0.069}$ &     $3.63^{+0.22}_{-0.13}$ &        $3670^{+170}_{-100}$ &     $-1.40^{+0.67}_{-1.00}$ &       $620^{+65}_{-53}$ &        $0.164^{+0.026}_{-0.016}$ &     $88.92^{+0.37}_{-0.29}$ &          $431^{+34}_{-32}$ &  $1.073^{+0.100}_{-0.088}$ \\
OGLE-TR-1092 &        $1.16^{+0.14}_{-0.16}$ &     $2.59^{+0.17}_{-0.30}$ &  $3.707^{+0.086}_{-0.140}$ &         $4804^{+250}_{-83}$ &      $0.20^{+0.20}_{-0.26}$ &     $2292^{+500}_{-15}$ &     $0.0493^{+0.0024}_{-0.0028}$ &        $78.8^{+2.3}_{-1.0}$ &         $1690^{+78}_{-51}$ &  $1.043^{+0.170}_{-0.046}$ \\
OGLE-TR-1093 &        $1.59^{+0.26}_{-0.18}$ &     $3.61^{+0.32}_{-0.29}$ &  $3.518^{+0.110}_{-0.079}$ &        $5220^{+590}_{-290}$ &      $0.17^{+0.27}_{-0.45}$ &    $2710^{+480}_{-260}$ &     $0.0317^{+0.0014}_{-0.0011}$ &        $57.4^{+3.4}_{-3.5}$ &       $2700^{+220}_{-120}$ &     $3.82^{+0.92}_{-1.10}$ \\
OGLE-TR-1094 &     $1.005^{+0.042}_{-0.053}$ &     $1.65^{+0.17}_{-0.22}$ &  $4.005^{+0.120}_{-0.087}$ &        $4990^{+170}_{-160}$ &      $0.35^{+0.14}_{-0.35}$ &    $1450^{+130}_{-140}$ &     $0.1095^{+0.0015}_{-0.0019}$ &     $89.19^{+0.58}_{-0.88}$ &          $930^{+43}_{-45}$ &     $1.62^{+0.16}_{-0.21}$ \\
OGLE-TR-1095 &        $1.83^{+4.40}_{-0.89}$ &     $2.04^{+0.21}_{-0.18}$ &     $4.11^{+0.46}_{-0.29}$ &     $9500^{+17000}_{-3100}$ &        $-1.8^{+1.6}_{-1.7}$ &  $5500^{+5900}_{-2100}$ &     $0.0205^{+0.0100}_{-0.0040}$ &       $69.9^{+14.0}_{-9.9}$ &     $4600^{+6000}_{-1200}$ &  $1.062^{+0.110}_{-0.092}$ \\
OGLE-TR-1096 &     $0.968^{+0.060}_{-0.110}$ &     $3.51^{+0.12}_{-0.15}$ &  $3.328^{+0.023}_{-0.023}$ &        $4560^{+220}_{-120}$ &      $0.05^{+0.34}_{-0.62}$ &    $2880^{+230}_{-150}$ &  $0.04410^{+0.00090}_{-0.00170}$ &        $88.4^{+1.1}_{-1.7}$ &         $1967^{+95}_{-56}$ &  $2.446^{+0.080}_{-0.089}$ \\
OGLE-TR-1097 &     $0.873^{+0.140}_{-0.089}$ &  $0.874^{+0.074}_{-0.072}$ &  $4.514^{+0.056}_{-0.064}$ &      $6200^{+1200}_{-1400}$ &        $-1.0^{+1.3}_{-2.5}$ &    $1280^{+840}_{-580}$ &     $0.0926^{+0.0046}_{-0.0032}$ &     $88.08^{+0.22}_{-0.24}$ &        $920^{+190}_{-220}$ &  $1.017^{+0.096}_{-0.093}$ \\
OGLE-TR-1098 &        $1.09^{+0.15}_{-0.16}$ &  $1.272^{+0.079}_{-0.082}$ &  $4.262^{+0.066}_{-0.072}$ &        $5960^{+470}_{-290}$ &      $0.16^{+0.25}_{-0.54}$ &    $2070^{+160}_{-150}$ &     $0.0656^{+0.0029}_{-0.0035}$ &     $86.06^{+0.46}_{-0.45}$ &        $1264^{+110}_{-64}$ &     $1.76^{+0.13}_{-0.14}$ \\
OGLE-TR-1099 &        $1.01^{+0.81}_{-0.40}$ &        $8.1^{+2.2}_{-1.8}$ &     $2.63^{+0.23}_{-0.20}$ &        $4460^{+180}_{-200}$ &     $-0.26^{+0.18}_{-0.47}$ &  $5500^{+1700}_{-1300}$ &        $0.107^{+0.017}_{-0.015}$ &        $70.9^{+3.8}_{-4.1}$ &       $1870^{+170}_{-150}$ &        $4.2^{+2.0}_{-1.1}$ \\
\end{supertabular}}
\medskip
\end{landscape}


\newpage

\begin{references}
\refitem{Alard, C., and Lupton, R.H.}{1998}{\ApJ}{503}{325}
\refitem{Bakos, G. and {Noyes}, R.W. and {Kov{\'a}cs}, G., \etal}{2004}{PASP}{116}{266}
\refitem{Baluev, R.}{2012}{\MNRAS}{422}{2372}
\refitem{Bouchy, F., Pont, F., Santos, N., \etal}{2004}{\AA}{421}{L13}
\refitem{Borucki, W. {Koch}, D., {Basri}, G., \etal}{2010}{Science}{327}{977}
\refitem{Chambers, K., {Magnier}, E., {Metcalfe}, N., \etal}{}{2016}{arXiv:1612.05560}{}
\refitem{Charbonneau, D., Brown, T., Latham, D., and Mayor, M.}{2000}{ApJ}{529}{L45}
\refitem{Choi, J., {Dotter}, A.,  {Conroy}, C., \etal}{\ApJ}{2016}{823}{102}
\refitem{Cumming, A., {Butler}, R.,{Marcy}, G, \etal}{2008}{\PASP}{120}{531}
\refitem{Eastman, J.,  {Siverd}, R., and {Gaudi}, S.}{2010}{\PASP}{122}{935}
\refitem{Eastman, J., {Rodriguez}, J., {Agol}, E., \etal }{}{2019}{arXiv:1907.09480}{}
\refitem{Gaia Collaboration, {Montegriffo}, P., {Bellazzini}, M., \etal}{2023a}{\AA}{674}{A33}
\refitem{Gaia Collaboration, {Vallenari}, A., {Brown}, A., \etal }{2023b}{\AA}{674}{A1}
\refitem{Gonzalez, O., Rejkuba, M., Zoccali, M., \etal}{2012}{\AA}{543}{A13}
\refitem{Gould, A., {Huber}, D., {Penny}, M., and {Stello}, D.}{2015}{Journal of Korean Astronomical Society}{48}{93}
\refitem{Guo, X., {Johnson}, J., {Mann}, A., \etal}{2017}{\ApJ}{838}{25}
\refitem{Hagey, S.R., {Edwards}, B., and {Boley}, A.C.}{2022}{AJ}{164}{220}
\refitem{Henry, G., Marcy, G., Butler, R., and Vogt, S.}{2000}{ApJ}{529}{L41}
\refitem{Hippke, M. and {Heller}, R.}{2019}{\AA}{623}{A39}
\refitem{Howard, A., {Marcy}, G., {Bryson}, S., \etal}{2012}{\ApJS}{201}{15}
\refitem{Konacki, M., {Torres}, G.,  {Jha}, S., and {Sasselov}, D.}{2003a}{Nature}{421}{507}
\refitem{Konacki, M., {Torres}, G., {Sasselov}, D., and {Jha}, S.,}{2003b}{\ApJ}{597}{1076}
\refitem{Kov{\'a}cs, G. and {Zucker}, S. and {Mazeh}, T.}{2002}{\AA}{391}{369}
\refitem{Mandel, K. and {Agol}, E.}{2002}{\ApJ}{580}{L171}
\refitem{Mayor, M., {Marmier}, M., {Lovis}, C., \etal}{2011}{}{arXiv:1109.2497}{}
\refitem{Mazeh, T., {Holczer}, T., and {Faigler}, S.}{2016}{\AA}{589}{A75}
\refitem{Melo, C., Santos, N., Pont, F., \etal}{2006}{\AA}{460}{251}
\refitem{Montet, B., {Yee}, J. and {Penny}, M.}{2017}{\PASP}{129}{044401}
\refitem{Morton, T.}{2012}{\ApJ}{761}{6}
\refitem{Morton, T.}{2016}{\ApJ}{822}{86}
\refitem{Nataf, D., {Gould}, A,. {Fouqu{\'e}}, P., \etal  }{2013}{\ApJ}{769}{88}
\refitem{Penny, M., {Gaudi}, B., and {Kerins} E. \etal} {2019}{\ApJS}{241}{3}
\refitem{Pollacco, D., {Skillen}, I., and {Collier Cameron}, A.}{2006}{PASP}{118}{1407}
\refitem{Pont, F., Bouchy, F., and Queloz, D., \etal }{2004}{\AA}{426}{L15}
\refitem{Pont, F.,  {Zucker}, S., {Queloz}, D.}{2006}{MNRAS}{373}{231}
\refitem{{Pont}, F., {Tamuz}, O., {Udalski}, A., \etal}{2008}{\AA}{487}{749}
\refitem{Ricker, G., {Winn}, J., {Vanderspek}, R., \etal }{2015}{Journal of Astronomical Telescopes, Instruments, and Systems}{1}{014003}
\refitem{Saydjari, A., {Schlafly}, E., {Lang}, D., \etal}{2023}{ApJ}{264}{L28}
\refitem{Sahu, K., Casertano, S., Bond, H., \etal }{2006}{Nature}{443}{534}
\refitem{Sahu, K., Casertano, S., Valenti, J., \etal }{2008}{ASPC}{398}{93}
\refitem{Soszy\'nski I., Pawlak M., Pietrukowicz P.,  \etal}{2016}{\Acta}{66}{405}
\refitem{Spergel, D., {Gehrels}, N., {Baltay}, C, \etal }{2015}{}{arXiv:1503.03757}{}
\refitem{Udalski, A., Paczy\'nski K., \.Zebru\'n, K., \etal }{2002a}{\Acta}{52}{1}
\refitem{Udalski, A., \.Zebru\'n, K., {Szyma\'nski}, M., \etal}{2002b}{\Acta}{52}{115}
\refitem{Udalski, A., Szewczyk, O., \.Zebru{\'n}, \etal}{2002c}{\Acta}{52}{317}
\refitem{Udalski, A., {Pietrzy\'nski}, G. , {Szyma\'nski}, M., \etal}{2003}{\Acta}{53}{133}
\refitem{Udalski, A., {Szyma\'nski}, M., Kubiak, M., \etal}{2004}{\Acta}{54}{313}
\refitem{Udalski, A., Pont, F., Naef, C., \etal}{2008}{\AA}{482}{749}
\refitem{Udalski, A., {Szyma{\'n}ski}, M., and M. {Szyma{\'n}ski}, G.}{2015}{\Acta}{65}{1}
\refitem{West, R.,{Pollacco}, D., {Wheatley}, P., \etal}{2016}{The Messenger}{165}{10}
\refitem{Wo\'zniak, P.R.}{2000}{\Acta}{50}{421}
\refitem{Zhu, W., and {Dong}, S.}{2021}{ARA\&A}{59}{291}
\end{references}
\end{document}